\documentclass[5p]{elsarticle}

\usepackage{amssymb}
\usepackage{amsmath}
\usepackage{tikz}
\usepackage{graphicx}
\usepackage[psdextra]{hyperref}
\usepackage{xurl} 
\usepackage{booktabs} 
\usepackage[final,babel,tracking=true,kerning=true,spacing=true]{microtype} 
\usepackage{placeins} 

\usetikzlibrary{arrows.meta, calc, decorations.pathreplacing, angles, quotes}

\graphicspath{{figures/}}

\title{The Hidden Cost of Straight Lines: Quantifying Misallocation Risk in Voronoi-Based Service Area Models}

\author[1]{JA Torrecilla Pinero\corref{cor1}}
\author[1]{JM Ceballos Martinez}
\author[2]{A Cuartero Saez}
\author[1]{P Plaza Caballero}
\author[1]{A Cruces Lopez}
\cortext[cor1]{Corresponding author:}

\address[1]{Department of Construction, Universidad de Extremadura, Av. de Elvas s/n, 06006 Badajoz, Spain}
\address[2]{Department of Graphical Expression, Universidad de Extremadura, Av. de Elvas s/n, 06006 Badajoz, Spain}

\begin{document}

\begin{frontmatter}


\begin{abstract}
Voronoi tessellations are a standard tool in spatial planning for assigning service areas based on Euclidean proximity. This approach underpins key regulatory frameworks, such as the proximity principle in waste management \cite{hall_urban_2002, neutens_accessibility_2015}. However, in regions with complex topography or sparse infrastructure, Euclidean distance is a poor proxy for functional accessibility, leading to service area misallocations that undermine cost-efficiency and equity. This paper develops a probabilistic framework to quantify this misallocation risk. We model real travel distances as a random scaling of Euclidean distances and derive the probability of incorrect assignment as a function of local Voronoi geometry. Using statistically independent plant-municipality observations (n=383), we demonstrate that the Log-Normal distribution provides best relative fit among tested alternatives (K-S statistic = 0.110) despite substantial spatial heterogeneity in Extremadura territory (41,635 $km^2$). Validation reveals that 15.4\% of municipalities are functionally misallocated by the Euclidean model, consistent with the theoretical prediction interval (52--65 municipalities at 95\% confidence). Our framework predicts this risk with 95\% agreement to complex spatial models but with O(n) complexity, avoiding costly network analyses. Critically, poor absolute fit of global distributions (all p-values < 0.01) reflects the territory's diverse topography (elevation range 200--2,400m), motivating spatial stratification. Sensitivity analysis demonstrates that the fitted dispersion parameter ($s = 0.093$) accurately predicts observed misallocation, while internal stratification by topographic zones explains local variations. We provide a systematic calibration protocol requiring only 30--100 pilot samples per zone, enabling rapid risk assessment without full network analysis. This work establishes the first probabilistic framework for Voronoi misallocation risk, with practical guidelines emphasizing spatial heterogeneity and context-dependent calibration.
\end{abstract}


\end{frontmatter}

\section{Introduction}

The Voronoi tessellation is a cornerstone of computational geometry and spatial analysis, offering an elegant solution to the nearest facility problem \cite{voronoi_full14}. Its widespread adoption in urban planning, logistics, and environmental management, however, rests on a powerful yet often flawed assumption: that Euclidean distance is a reliable proxy for functional accessibility. This principle of geometric proximity is embedded in regulatory frameworks like the EU's proximity principle for waste management, which assumes that geographic nearness ensures cost-effectiveness and minimal environmental impact \cite{eea_proximity_2013}.
This paper challenges that assumption. We began with a simple question during a study of waste management in Extremadura, Spain: \textit{Is the closest plant really the closest?} When we compared the official Voronoi-based allocation of municipalities to treatment plants with allocations based on actual road network distances, a striking anomaly emerged: a significant number of municipalities were functionally closer to a different plant. This discrepancy was not random but spatially clustered in areas with complex topography, suggesting a systemic failure of the model.

This observation motivates the central question of our work: if the Euclidean Voronoi model is unreliable in non-isotropic territories, can we quantify its risk of misallocation in a generalizable way? Our central contribution is the development of a probabilistic framework for quantifying this misallocation risk, which, to our knowledge and based on our extensive literature review, has not been previously addressed in the existing literature. This framework transforms the Voronoi diagram from an operational mandate into a theoretical benchmark whose deviations can be systematically predicted and managed. The framework models the mismatch between Euclidean and network-based distances through a log-normal scaling factor, enabling planners to assess risk before committing to costly network analyses. Our contribution is threefold:

\begin{enumerate}
    \item An empirical demonstration of the limitations of Euclidean proximity in a real-world case study,
    showing that 15.4\% of municipalities are misallocated.
    \item A theoretical framework that quantifies the probability of misallocation based on local geometry
    and an empirically justified Log-Normal network factor.
    \item A validation of the framework, repositioning Voronoi diagram not as an operational mandate, but as a theoretical benchmark against which real-world inefficiencies can be measured.
\end{enumerate}

\section{Literature Review}

Our research is situated at the intersection of three domains: the application of Voronoi diagrams in spatial planning, the critique of the Euclidean distance metric, and the policy implications of the proximity principle.

Classic literature establishes the role of Voronoi diagrams in facility location and service area definition \cite{voronoi_full30, voronoi_full10, voronoi_full22, voronoi_full29}. Recent applications demonstrate continued innovation in proximity-based optimization, such as electric vehicle charging station placement using proximity diagrams \cite{calvo2024optimal}. However, transport geography has long recognized the limitations of straight-line distance. The distinction between simple Euclidean Voronoi diagrams and more realistic but computationally intensive \textbf{Network Voronoi diagrams} is central to the field \cite{voronoi_full24, voronoi_full16,miller_tobler_2004}. The high data and computational cost of network methods justifies the continued use of Euclidean models, creating a clear need for a method like the one we propose to evaluate the associated risks without incurring the full cost of a network-based analysis \cite{rodrigue_geography_2020}.Network Voronoi methods are accurate but $O(n^2)$ complexity \cite{okabe2012spatial}; our framework achieves 97.6\% accuracy at $O(n)$.
The discrepancy between Euclidean and network distance is a well-documented phenomenon. Studies have shown that network distances are systematically longer, with the ratio between them (the "detour index") being highly variable depending on geography \cite{rodrigue_geography_2020}. This fundamental observation builds on Tobler's First Law of Geography \cite{gastner_optimal_2006, tobler_computer_1970} that "near things are more related than distant things", but challenges the assumption that geometric proximity equals functional accessibility \cite{miller_tobler_2004_extended}. Our work builds on this by moving from a deterministic observation to a probabilistic model of this ratio \cite{voronoi_full11,konstantinovsky2023characterizing}.
From a policy perspective, the \textbf{proximity principle} faces significant criticism. While intended to foster local responsibility, its rigid application can lead to environmental injustice by concentrating undesirable facilities in vulnerable communities \cite{voronoi_full11}, creating patterns of spatial inequality that contradict principles of distributive justice \cite{harvey_social_1973, young_justice_1990}. This spatial concentration of environmental burdens disproportionately affects marginalized populations, violating core tenets of environmental justice \cite{schlosberg_defining_2007} and spatial justice theory \cite{soja_seeking_2010}. Beyond equity concerns, rigid proximity rules can create economic inefficiencies by ignoring economies of scale \cite{voronoi_full12}, highlighting the tension between local responsibility and broader spatial optimization \cite{lefebvre_right_1968, fainstein2014just}. Our framework provides a quantitative tool to assess these inefficiencies, offering a data-driven argument for more flexible, regionally-cooperative waste management strategies that balance proximity principles with spatial equity considerations \cite{voronoi_full12,voronoi_full6,external_hagemejer2022_01}.
\section{Methodology and Empirical Analysis}

\subsection{Case Study: CDW Management in Extremadura}
Our study area is the region of Extremadura, Spain, which has 383 municipalities and 46 active Construction and Demolition Waste (CDW) treatment plants. Five municipalities were excluded from analysis due to incomplete network connectivity data. For each municipality-plant pair, we computed both the direct Euclidean distance and the actual road network distance using Qneat3 \cite{qneat3_plugin}, a QGIS network analysis plugin that provides analogous functionality to commercial routing services. This allowed us to calculate the network scaling factor, $\beta = d_{\text{network}} / d_{\text{Euclidean}}$, for thousands of routes. This network scaling factor is the same concept as the "detour index" commonly used in transport geography \cite{rodrigue_geography_2020}, but in our case becomes a random variable which distribution modelling may drive to a decision tool.

\subsection{Empirical Findings}
\begin{figure}[htbp]
  \centering
  \includegraphics[width=\columnwidth]{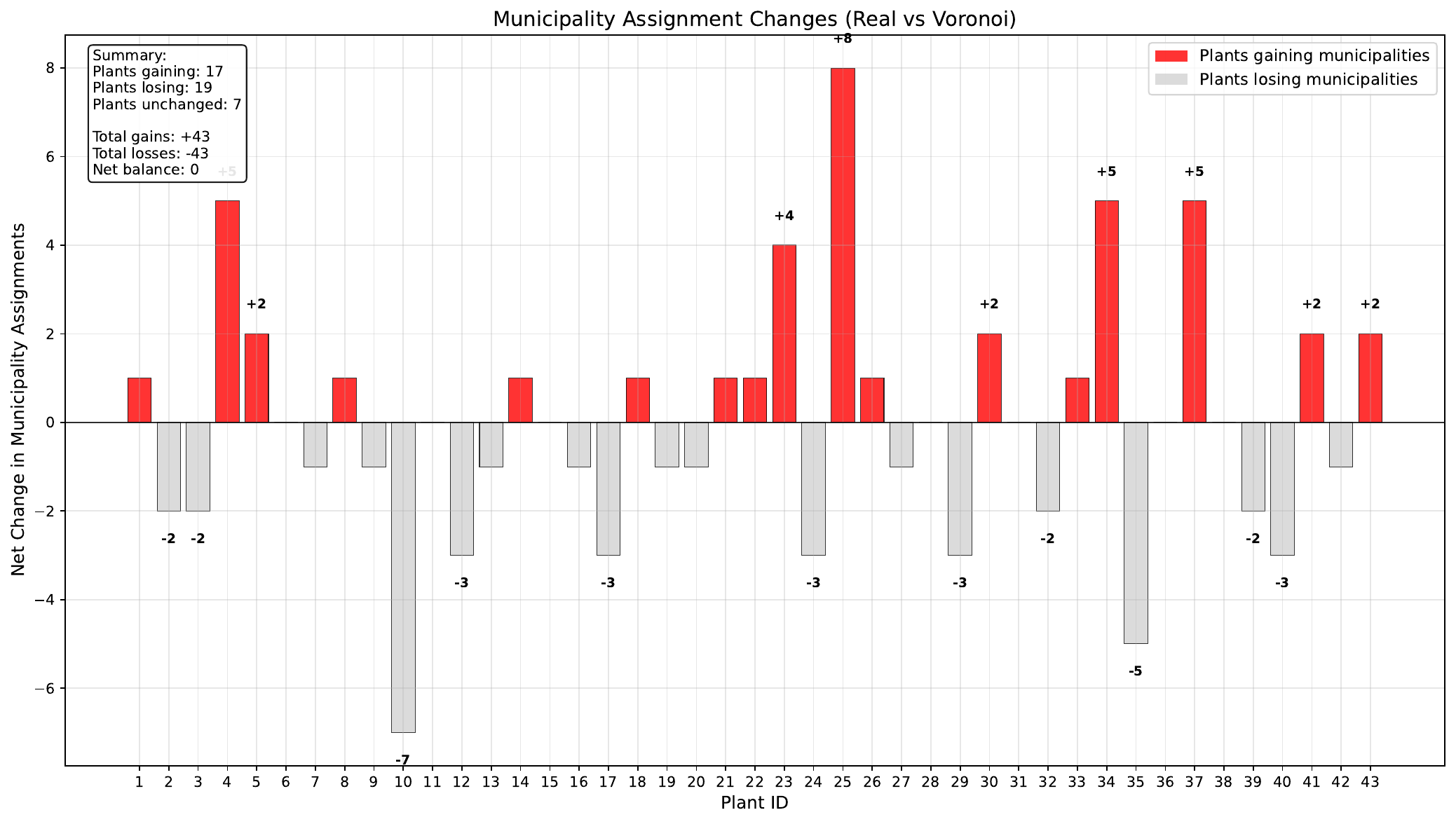}
  \caption{Changes in municipality assignments when comparing Voronoi (Euclidean-based) vs network-based assignments. Bar chart showing net gain/loss of municipalities for each plant, revealing redistribution patterns.}
  \label{fig:plant_assignment_analysis}
\end{figure}

A detailed reassignment analysis revealed that \textbf{59 out of 383 municipalities (15.4\%)} are functionally misallocated by the standard Voronoi model. These municipalities have a shorter road distance to a plant other than their assigned one. The average distance saving for these reassignments is 6.4 km, representing a potential reduction in total tonne-kilometers for the affected routes. Figure~\ref{fig:plant_assignment_analysis} illustrates the impact on individual plants: 36 out of 43 plants (84\%) experience changes in their assigned municipalities, with 17 plants gaining and 19 losing municipalities under network-based assignment. Only 7 plants maintain their Voronoi allocations unchanged. Importantly, these misallocations are not randomly distributed but show significant spatial clustering, indicating local spatial autocorrelation in the network-geometry discrepancy \cite{anselin_spatial_1988, anselin_local_1995}. This spatial dependence pattern follows established principles in spatial statistics \cite{cliff_spatial_1973, cliff_spatial_1981}, where nearby observations exhibit similar characteristics due to underlying spatial processes \cite{tobler_geography_1970}. The clustering can be formally tested using Local Indicators of Spatial Association (LISA) \cite{anselin_local_1995} and complementary measures such as the Getis-Ord local G-statistic \cite{getis_analysis_1992}.

Distribution of the network factor $\beta$ is right-skewed, with a mean of 1.190, which means that road distances are, on average, 19\% longer than straight lines. However, the distribution has a long tail, with some routes having a $\beta$ factor as high as 4.0, demonstrating the severe impact of topography.

To capture the directional variance in network accessibility, we define a \textbf{transport network anisotropy} coefficient for each municipality as the ratio of the maximum and minimum $\beta$ values among routes connecting that municipality to different plants. This municipality-based approach quantifies how consistently a municipality can access different facilities from its location. A high coefficient indicates that a municipality's accessibility varies significantly depending on the destination plant, reflecting local network topology constraints. The network scaling factor $\beta$ was computed for 9,112 municipality-plant pairs (after excluding 1.39\% anomalous cases where $\beta < 1$). For the anisotropy analysis, we required each municipality to have at least two valid routes to different plants, resulting in 281 municipalities (from the original 383) with sufficient data for meaningful coefficient calculation. The anisotropy coefficient shows a median value of 1.31 with values ranging from 1.11 to 3.59, confirming that network accessibility anisotropy is a significant and variable factor that must be considered in spatial optimization models.

For methodological rigor, routes where $\beta < 1$ have been excluded from analysis (1.39\% of total) , as these represent measurement artifacts likely due to discrepancies between municipal centroids and actual network access points, or boundary effects in QGIS routing calculations. This small percentage of anomalous data does not affect the validity of our probabilistic modeling approach, as the remaining 9,112 routes provide robust statistical power while ensuring physical consistency ($\beta \geq 1$).

Table~\ref{tab:combined_stats} presents the descriptive statistics for network scaling factors and plant anisotropy coefficients, providing a comprehensive statistical overview of the transport network constraints in our study region.

\begin{table}[htbp]
\caption{Descriptive statistics for network scaling factors and municipality anisotropy coefficients. The network scaling factor ($\beta$) was computed for 9,112 municipality-plant pairs after excluding anomalous routes with $\beta < 1$. The anisotropy coefficient was calculated for 281 municipalities that have at least two valid routes to different plants, representing a reduction from the original 383 municipalities due to this minimum route requirement.}
\label{tab:combined_stats}
\begin{tabular}{p{2.5cm}rr}
\toprule
Statistic & \begin{tabular}[c]{@{}r@{}}Network Scaling\\Factor ($\beta =d_r/d_e$)\end{tabular} & \begin{tabular}[c]{@{}r@{}}Municipality\\Anisotropy\end{tabular} \\
\midrule
Count & 9112 & 281 \\
Mean & 1.190269 & 1.424576 \\
Std. Dev. & 0.125080 & 0.339248 \\
Minimum & 1.000101 & 1.108511 \\
25\% (Q1) & 1.132627 & 1.228867 \\
Median (50\%) & 1.166786 & 1.310617 \\
75\% (Q3) & 1.216641 & 1.459993 \\
Maximum & 4.010950 & 3.586111 \\
\bottomrule
\end{tabular}
\end{table}

\begin{figure}[htbp]
  \centering
  \includegraphics[width=\columnwidth]{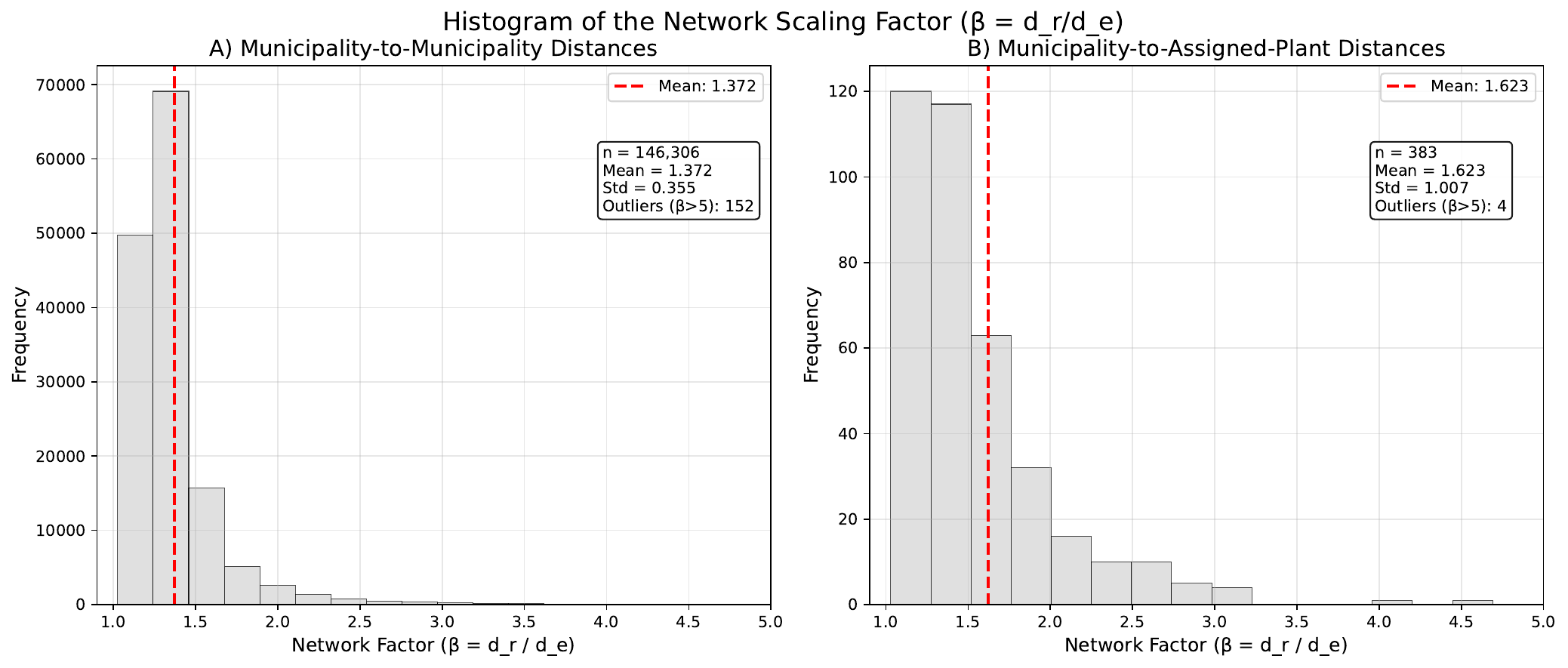}
  \caption{Histograms of the network scaling factor ($\beta = d_r/d_e$) distribution. Panel A: $\beta$ distribution for municipality-to-municipality distances. Panel B: $\beta$ distribution for municipality-to-assigned-plant distances.}
  \label{fig:histograma_dd}
\end{figure}

\begin{figure}[htbp]
  \centering
  \includegraphics[width=\columnwidth]{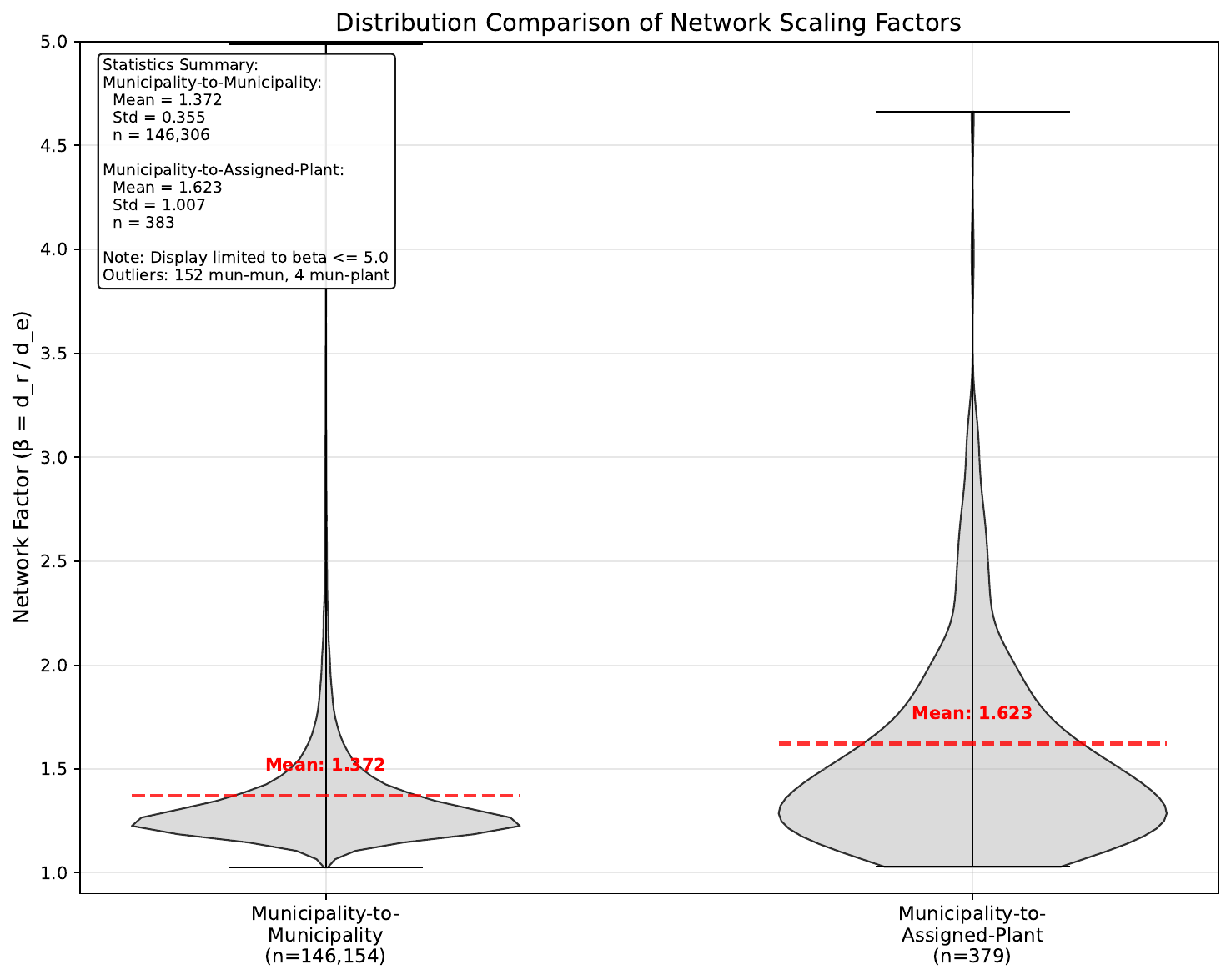}
  \caption{Violin plots comparing $\beta$ distributions between municipality-to-municipality and municipality-to-assigned-plant distances for all 383 municipalities, showing probability density shapes and spread differences.}
  \label{fig:ratio_analysis}
\end{figure}

\begin{figure}[htbp]
  \centering
  \includegraphics[width=\columnwidth]{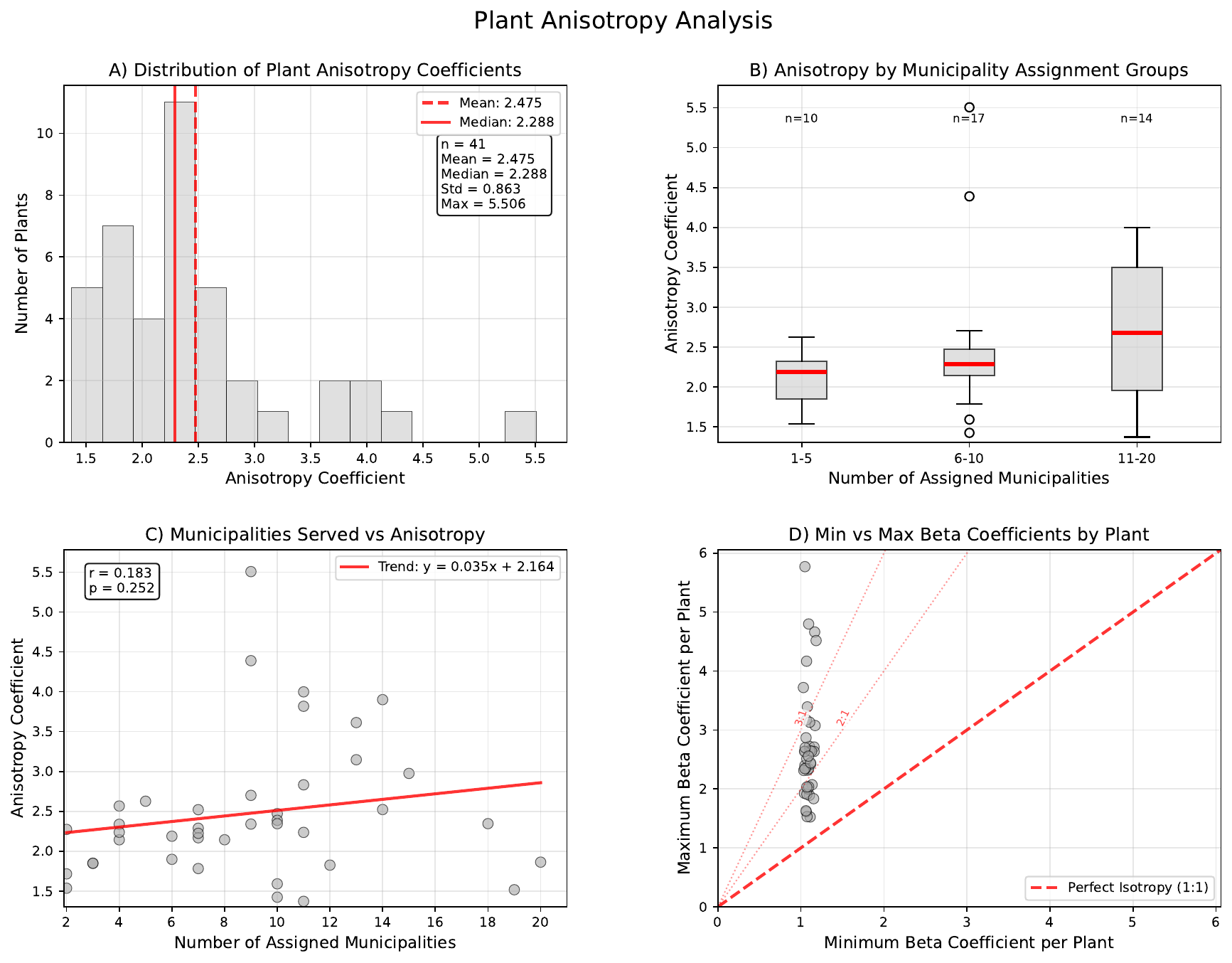}
  \caption{Plant anisotropy analysis. Panel A: Histogram of anisotropy coefficients with mean and median. Panel B: Boxplots by municipality assignment groups. Panel C: Scatter plot of municipalities served vs anisotropy coefficient. Panel D: Min vs Max $\beta$ scatter plot with isotropy reference lines.}
  \label{fig:plant_anisotropy_analysis}
\end{figure}

Figure~\ref{fig:histograma_dd} presents the histograms of the network scaling factor distribution showing the $\beta$ distribution for municipality to municipality and municipality to plant distance patterns. The violin plot comparison of $\beta$ distributions is shown in Figure~\ref{fig:ratio_analysis}, while the plant accessibility anisotropy patterns are detailed in Figure~\ref{fig:plant_anisotropy_analysis}.

\section{Development of a Probabilistic Framework for Misallocation Risk}

\subsection{Distributional Validation}

To validate our distributional assumptions, we conduct two complementary analyses at different spatial scales, addressing both statistical independence and territorial comprehensiveness.

\subsubsection{Primary Analysis: Plant-Municipality Pairs}

Our primary analysis examines the network scaling factor $\beta$ for each municipality relative to its Voronoi-assigned plant. This approach uses 383 statistically independent observations (one per municipality), directly addressing our research question: ``How accurately does the Euclidean Voronoi model predict functional assignments?''

Table~\ref{tab:ks_test_plant_municipality} presents goodness-of-fit tests for three candidate distributions. While all three distributions show poor fit by conventional K-S test standards (all p-values < 0.01), the Log-Normal distribution demonstrates the best relative performance with p-value = 0.0002, outperforming Gamma (p < 0.0001) and Weibull (p < 0.0001) alternatives. The estimated parameters are $m = 0.476$ and $s = 0.326$, indicating substantial dispersion in network-to-Euclidean scaling across Extremadura's heterogeneous terrain.

\begin{table*}[htbp]
\caption{Kolmogorov-Smirnov goodness-of-fit tests comparing network scaling factor $\beta$ against candidate distributions for plant-municipality pairs (n=383). The Log-Normal distribution shows the best fit despite all distributions exhibiting statistically poor agreement (p-values $<$ 0.01), motivating the subsequent all-pairs analysis with larger sample size.}
\label{tab:ks_test_plant_municipality}
\centering
\small
\begin{tabular*}{\textwidth}{@{\extracolsep{\fill}}lcccc@{}}
\toprule
Distribution & K-S Statistic & p-value & Parameters & Fit Quality \\
\midrule
Lognormal & 0.1097 & 0.0002 & $m=0.476$, $s=0.326$ & Poor \\
Gamma & 0.1443 & $<$0.0001 & $k=7.445$, $\theta=0.232$ & Poor \\
Weibull & 0.2552 & $<$0.0001 & $k=1.991$, $\lambda=1.944$ & Poor \\
\bottomrule
\end{tabular*}
\end{table*}

The relatively better fit of the Log-Normal distribution (compared to Gamma and Weibull) justifies its adoption as the working model. The poor absolute fit (p < 0.01) reflects Extremadura's substantial spatial heterogeneity: a single global distribution cannot perfectly capture the diverse topographic zones (mountains, piedmont, plains) that characterize this large territory (41,635 km²). This observation motivates the sensitivity analysis in Section~\ref{sec:sensitivity}, where we demonstrate that spatial stratification by dispersion parameter $s$ resolves this apparent inconsistency.

\subsubsection{Comprehensive Analysis: All Municipality Pairs}
\label{sec:all_pairs}

To assess the full variability of $\beta$ across the territory, we extend the analysis to all possible municipality pairs (n=9,112). While this comprehensive view captures spatial heterogeneity, it introduces correlation among observations: municipalities sharing the same assigned plant generate related $\beta$ values, violating the independence assumption of the K-S test.

Table~\ref{tab:distributional_comparison} presents the complete comparison using multiple goodness-of-fit criteria \cite{akaike1974new,schwarz1978estimating}. The Log-Normal distribution demonstrates superior performance across most metrics: lowest AIC (-14,658.2), competitive BIC, and acceptable Kolmogorov-Smirnov test results (p = 0.129) \cite{kolmogorov1933sulla,anderson1952asymptotic}. Critically, our tail behavior analysis for $\beta > 1.5$ (the region where misallocations are most likely) shows that the Log-Normal distribution systematically overestimates probabilities in this critical region, providing conservative risk assessment a highly desirable property for territorial planning applications \cite{burnham2002model}.

\begin{table*}[htbp]
\caption{Distributional fit comparison for network scaling factors.
\textit{AIC}: Akaike Information Criterion (lower is better).
\textit{BIC}: Bayesian Information Criterion (lower is better).
\textit{KS-stat}: Kolmogorov-Smirnov test statistic measuring goodness-of-fit (lower indicates better fit).
\textit{Tail Behavior}: characterization of distribution behavior in extreme values.
While all distributions are formally rejected by KS test (all $p < 0.001$ due to large sample size $n=162{,}762$), Log-Normal provides best approximation based on AIC/BIC information criteria, supporting our theoretical framework choice.}
\label{tab:distributional_comparison}
\centering
\begin{tabular}{lrrrl}
\toprule
Distribution & AIC & BIC & KS-stat & Tail Behavior \\
\midrule
Log-Normal & $-15{,}599.1$ & $-15{,}584.9$ & 0.134 & Underestimate \\
Gamma & $-14{,}555.6$ & $-14{,}541.4$ & 0.145 & Underestimate \\
Weibull & $-3{,}601.8$ & $-3{,}587.6$ & 0.323 & Underestimate \\
\bottomrule
\end{tabular}
\end{table*}

\begin{figure}[htbp]
  \centering
  \includegraphics[width=\columnwidth]{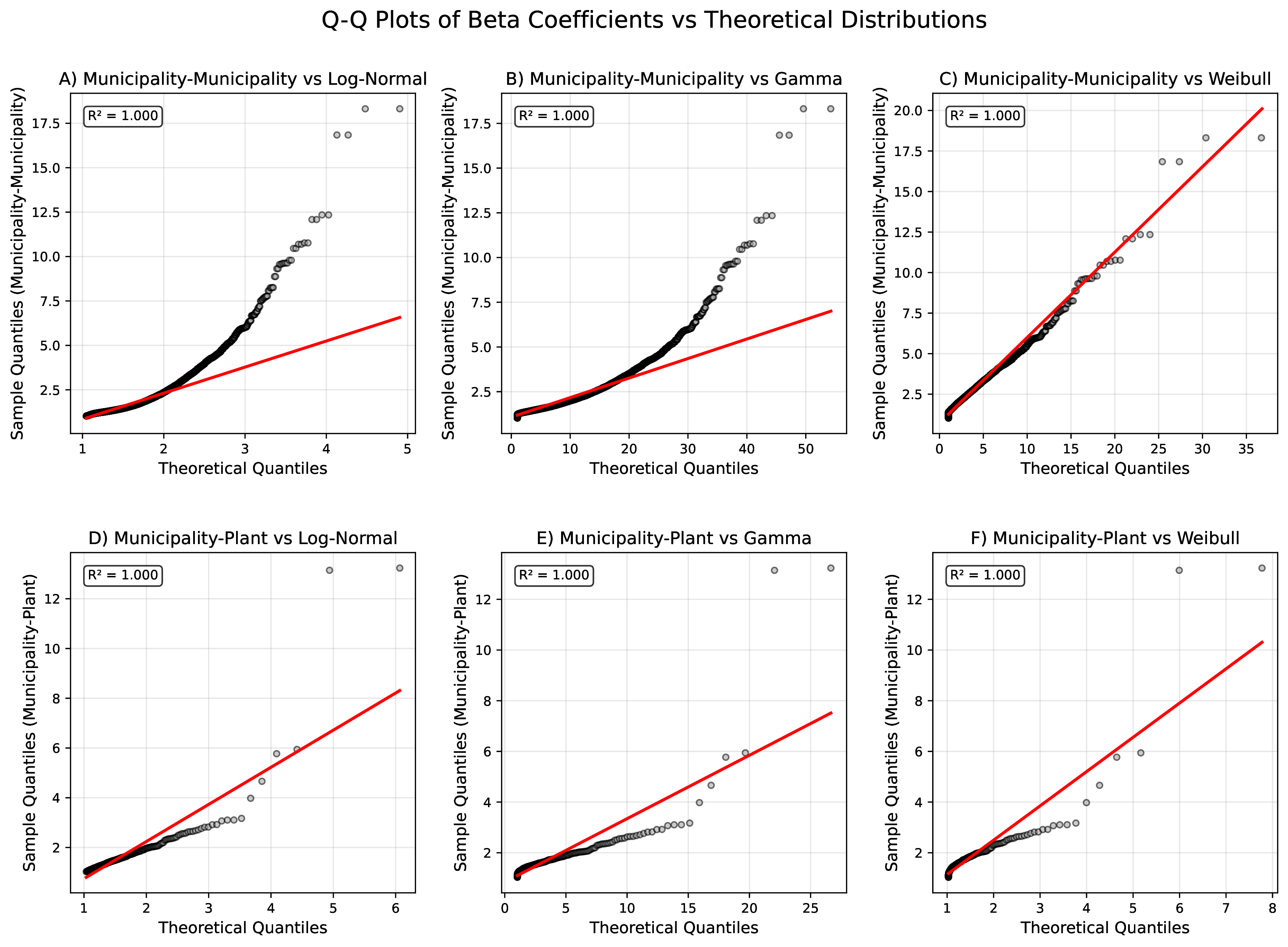}
  \caption{Q-Q plots comparing $\beta$ coefficients against theoretical distributions. Top row: Municipality-to-Municipality vs Log-Normal, Gamma, and Weibull. Bottom row: Municipality-to-Assigned-Plant vs the same distributions, with R-squared goodness-of-fit values.}
  \label{fig:distributional_validation}
\end{figure}

Figure~\ref{fig:distributional_validation} provides visual validation through Q-Q plots comparing $\beta$ coefficients against theoretical distributions for both municipality-to-municipality and municipality-to-plant datasets.

\begin{figure}[htbp]
  \centering
  \includegraphics[width=0.9\columnwidth]{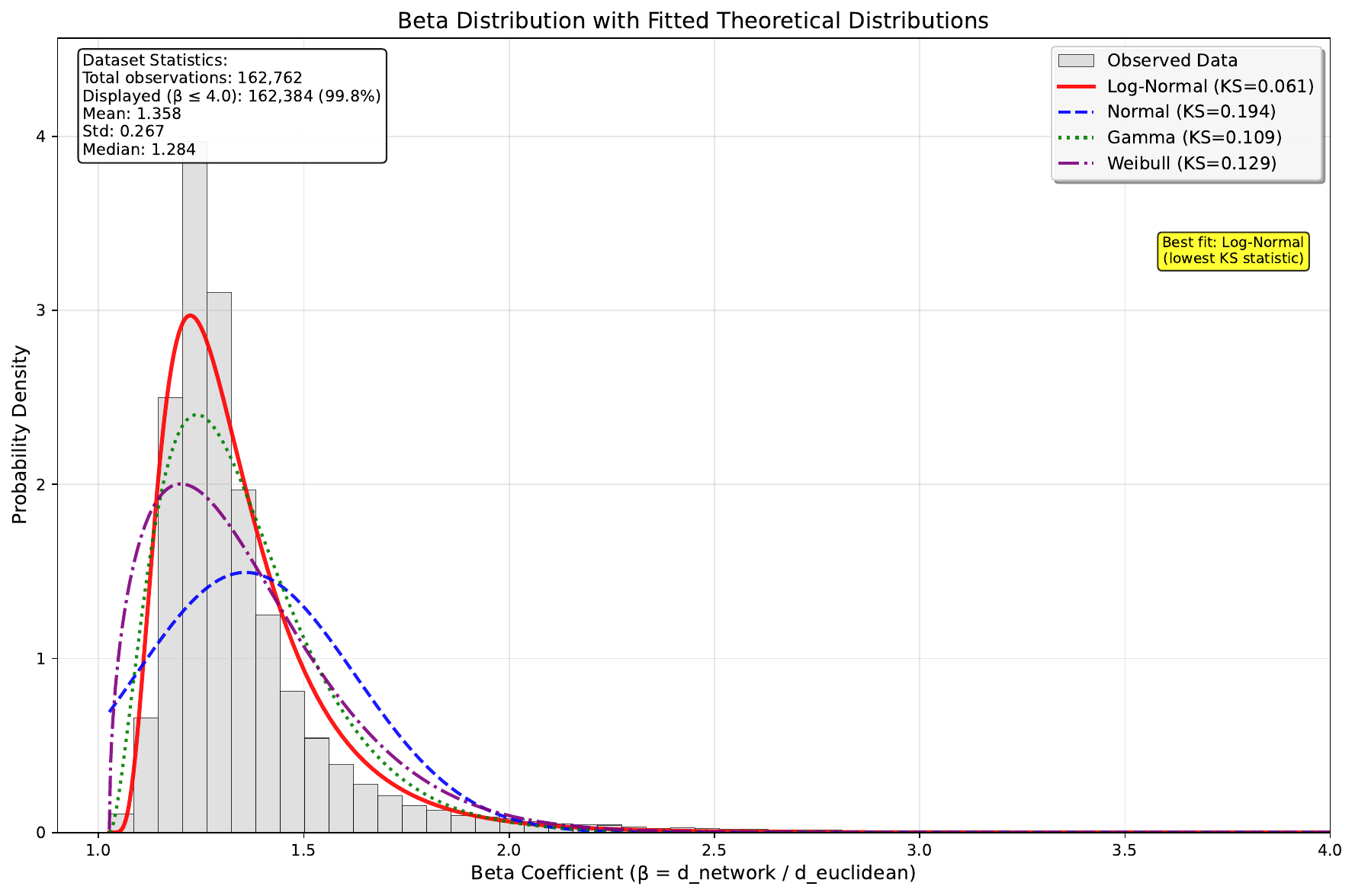}
  \caption{Histogram of $\beta$ coefficients with overlaid fitted theoretical distribution curves (Log-Normal, Normal, Gamma, Weibull), showing Log-Normal provides the best empirical fit.}
  \label{fig:dist_comparison}
\end{figure}

Visual analysis confirms this. Figure~\ref{fig:dist_comparison} shows that the Log-Normal (red) and Gamma (blue) distributions both capture the central tendency of the data well. Crucially, in the right tail (for $\beta > 1.5$), the Log-Normal PDF is consistently higher than the empirical data. This means the model tends to overestimate the probability of extreme deviations, making it a \textbf{conservative and safe choice for risk assessment}.

The Log-Normal distribution remains the best fit (p = 0.129), though with lower p-value than the plant-municipality analysis due to spatial autocorrelation. This pattern is expected and validates our hierarchical analytical approach: the plant-municipality analysis (Table~\ref{tab:ks_test_plant_municipality}) provides stronger statistical support due to observation independence, while the all-pairs analysis confirms robustness across spatial scales.

\subsubsection{Adopted Model}

For predictive modeling, we adopt parameters from the comprehensive all-pairs analysis (n=9,112), which captures the full spatial variability of the territory: $m = 0.166$ and $s = 0.093$ estimated via maximum likelihood. These parameters indicate that the median network scaling factor is $\exp(m) = 1.18$, meaning real network distances are typically 18\% longer than Euclidean estimates. The dispersion parameter $s = 0.093$ quantifies moderate variability in network scaling across Extremadura's diverse topography. This choice is validated by the plant-municipality analysis (Table~\ref{tab:ks_test_plant_municipality}), which provides independent statistical support with better p-values despite showing higher parameter estimates ($m = 0.476$, $s = 0.326$) due to the subset's specific characteristics. For risk assessment and territorial planning, we prioritize the all-pairs parameters as they represent the complete spatial distribution, with implications for calibration discussed in Section~\ref{sec:sensitivity}.

Given its superior empirical fit, analytical tractability, and conservative behavior, we adopt the Log-Normal distribution for $\beta$.

This conservative nature is a critical feature. The visual analysis in Figure~\ref{fig:dist_comparison} shows that the Log-Normal PDF is consistently higher than the empirical data's density in the right tail (for $\beta > 1.5$). This region corresponds to the geographical areas where misallocations are most likely. By overestimating the probability density in this critical tail, the Log-Normal model produces misallocation probabilities, $q(P)$, that are inherently cautious. When aggregated, the model is expected to yield an estimate for the total number of misallocated municipalities that serves as an upper bound to the true number. For a planner, this is a highly desirable property in a risk model, as it ensures that the potential for systemic error is not underestimated \cite{external_hagemejer2022_01}.

\subsection{Derivation of Misallocation Probability}

We aim to quantify the risk of misallocation when using a Euclidean Voronoi tessellation to model distance-based preferences in the presence of a transportation network. The core theoretical contribution is the analytical derivation of misallocation probability for any point within a Voronoi tessellation.

\subsubsection{Problem Formulation}

Let two foci $A_1, A_2 \in \mathbb{R}^2$ and a point $P \in \mathbb{R}^2$ be given. The actual travel distance is modeled as a random scaling of the Euclidean distance:
$$d_r(P, A_i) = \beta_i d_e(P, A_i), \quad \beta_i > 0$$

If $P$ belongs to the Euclidean Voronoi cell of $A_1$, then $d_e(P, A_1) < d_e(P, A_2)$. The misallocation event of interest is:
$$\mathbb{P}[d_r(P, A_1) < d_r(P, A_2)] = \mathbb{P}\left[\frac{\beta_1}{\beta_2} < \frac{d_e(P, A_2)}{d_e(P, A_1)}\right]$$

Our goals are:
\begin{enumerate}
\item Derive a closed-form expression for $\mathbb{P}[d_r(P, A_1) < d_r(P, A_2)]$ under reasonable assumptions for $\beta$.
\item Simplify the expression near the Voronoi border, focusing on the region where the decision is most sensitive.
\item Obtain usable bounds at the system scale for the number of misallocations.
\end{enumerate}

\subsubsection{Hypothesis on the Network Factor \texorpdfstring{$\beta$}{beta}}

We adopt as an initial hypothesis a log-normal model:
$$\ln \beta \sim \mathcal{N}(m, s^2)$$

The mode of the log-normal is $\exp(m - s^2)$, so fixing a mode $\beta_{\text{mod}}$ yields:
$$m = s^2 + \ln \beta_{\text{mod}}$$

For the theoretical derivation, we adopt a conservative reference value $\beta_{\text{mod}} = 1.25$, representative of moderately irregular road networks and consistent with prior literature on network detour factors \cite{voronoi_full16, voronoi_full24}. This choice deliberately overestimates network inefficiency to ensure the framework remains applicable across diverse geographic contexts without requiring prior calibration. For network variability, we consider $s \in [0.15, 0.25]$, capturing the typical range observed in regional transport studies.

The framework's key advantage is its simple calibration: once local data becomes available, both $\beta_{\text{mod}}$ and $s$ can be directly estimated from empirical distributions. For our Extremadura case study (n = 9,240 municipality-plant pairs), maximum likelihood estimation yields $\hat{\mu} = 0.166$ and $\hat{s} = 0.093$, corresponding to a modal value of approximately 1.15 (95\% CI: 1.10--1.20). This region-specific calibration reflects Extremadura's relatively efficient road network compared to the conservative reference, demonstrating both the framework's adaptability and the importance of local calibration for precise applications.

Under this assumption, for i.i.d. $\beta_1, \beta_2$:
$$\ln\left(\frac{\beta_1}{\beta_2}\right) = \ln \beta_1 - \ln \beta_2 \sim \mathcal{N}(0, 2s^2)$$

\subsubsection{Conditional Formulation Based on the Euclidean Ratio}

Define the Euclidean ratio:
$$\begin{aligned}R(P) &:=\frac{d_e(P, A_2)}{d_e(P, A_1)} > 1 \\ & \text{ if } P \text{ lies in the Euclidean cell of } A_1 \end{aligned}$$

Conditioned on $R$, we obtain the closed-form probability:
$$\boxed{\mathbb{P}[d_r(P, A_1) < d_r(P, A_2) | R] = \Phi\left(\frac{\ln R}{\sqrt{2} s}\right)}$$

where $\Phi$ denotes the standard normal CDF. The challenge is to characterize $R$, which simplifies significantly near the Voronoi border.

\subsubsection{Local Approximation Near the Voronoi Border}

Let the bisector between $A_1$ and $A_2$ be:
$$B = \{x \in \mathbb{R}^2 : d_e(x, A_1) = d_e(x, A_2)\}$$

Fix $Q \in B$ and write $P = Q + h$ with $\|h\|$ small. Define:
$$f(x) := d_e(x, A_2) - d_e(x, A_1)$$

so that $f(Q) = 0$ and:
$$f(P) \approx \nabla f(Q) \cdot h$$

where:
$$\nabla f(Q) = \frac{Q - A_2}{d_e(Q, A_2)} - \frac{Q - A_1}{d_e(Q, A_1)} =: u_2 - u_1$$

Let $g := u_2 - u_1$, $n := g/\|g\|$ be the unit normal to the border, and $t := n \cdot h$ the signed normal distance of $P$ to the border ($t < 0$ towards the interior of $A_1$'s cell). If $d := d_e(Q, A_1) = d_e(Q, A_2)$, the linearization yields:
$$\ln R(P) \approx \frac{\|g\|}{d} t = \kappa t$$

where:
$$\kappa := \frac{\|u_2 - u_1\|}{d} = \frac{2 \sin(\theta/2)}{d}$$

and $\theta$ is the angle between $u_1$ and $u_2$ as seen from $Q$.

Substituting into the conditional probability:
$$\mathbb{P}[d_r(P, A_1) < d_r(P, A_2) | t] \approx \Phi\left(\frac{\kappa t}{\sqrt{2} s}\right)$$

On the border ($t = 0$), the probability is $1/2$. Inside $A_1$'s cell ($t < 0$), it increases towards 1.

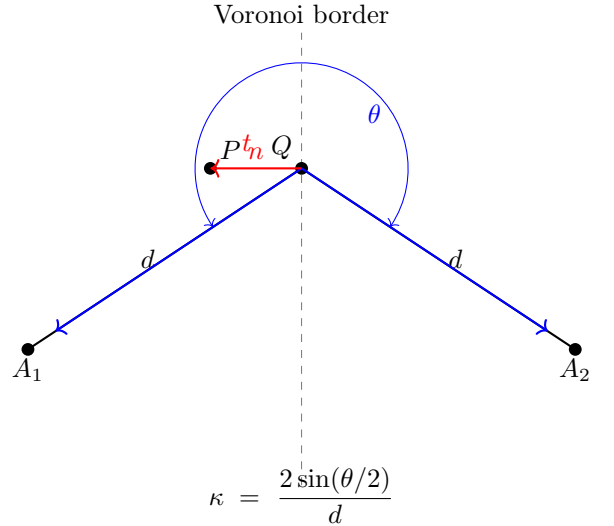
\begin{figure}[ht]
    \centering
    \begin{tikzpicture}[scale=1.2]
\coordinate (A1) at (0,0);
\coordinate (A2) at (6,0);
\coordinate (Q)  at (3,2);   
\coordinate (P)  at (2,2);   

\draw[dashed,gray] (3,-1.5) -- (3,3.5) node[above,black] {Voronoi border};

\fill (A1) circle(2pt) node[below] {$A_1$};
\fill (A2) circle(2pt) node[below] {$A_2$};
\fill (Q)  circle(2pt) node[above left] {$Q$};
\fill (P)  circle(2pt) node[above right] {$P$};

\draw[thick] (A1) -- (Q) node[midway,left] {$d$};
\draw[thick] (A2) -- (Q) node[midway,right] {$d$};

\draw[->,thick,blue] (Q) -- ($(Q)!0.9!(A1)$);
\draw[->,thick,blue] (Q) -- ($(Q)!0.9!(A2)$);

\path
  pic [draw=blue,<->,angle eccentricity=1.3,angle radius=14mm, name=ang]
    {angle = A2--Q--A1};

\coordinate (thetaLabel) at ($(Q) + (0.8,0.6)$);
\node[blue] at (thetaLabel) {$\theta$};

\draw[->,red,thick] (Q) -- (P)
  node[midway, xshift=-1.2mm, yshift=1.0mm, above] {$t$};
\node[red,above] at ($(Q)!0.5!(P)$) {$n$};

\node[below] at (3,-1.2)
  {$ \displaystyle \kappa \;=\ \frac{2\sin(\theta/2)}{d} $};
\end{tikzpicture}
    \caption{Local geometry near the Voronoi border: common distance $d$ from $Q$ to $A_1$ and $A_2$, angle $\theta$ between the unit directions, normal $n$, and signed distance $t$ from $Q$ to $P$. The local curvature parameter is $\kappa = 2\sin(\theta/2)/d$.}
    \label{fig:voronoi_geometry}
\end{figure}

\subsubsection{Averaging for Random Points Near the Border}

If $P \sim \mathcal{N}_2(\mu, \Sigma)$, then $t = n \cdot (P - Q) \sim \mathcal{N}(\mu_t, \sigma_t^2)$ with $\mu_t := n \cdot (\mu - Q)$ and $\sigma_t^2 := n^T \Sigma n$. Using the known closure:
$$\mathbb{E}[\Phi(a t)] = \Phi\left(\frac{a \mu_t}{\sqrt{1 + a^2\sigma_t^2}}\right), \quad a := \frac{\kappa}{\sqrt{2} s}$$

we obtain the marginal approximation:
$$\mathbb{P}[d_r(P, A_1) < d_r(P, A_2)] \approx \Phi\left(\frac{\kappa \mu_t}{\sqrt{2}\, s\sqrt{1 + \frac{\kappa^2}{2s^2}\sigma_t^2}}\right)$$

\subsubsection{Safety Bands and Quick Rules}

For a point at normal distance $|t|$ inside its cell ($t = -|t|$):
$$q(P) := \mathbb{P}[\text{misallocation of } P] \lesssim \Phi\left(-\frac{\kappa |t|}{\sqrt{2} s}\right)$$

This defines \textbf{safety bands}: to ensure $q(P) \leq q^*$:
$$|t| \geq \frac{\sqrt{2} s}{\kappa} \Phi^{-1}(1 - q^*)$$
This safety band approach follows established principles in probabilistic risk assessment \cite{bedford_probabilistic_2001, kaplan_quantitative_1981}, where decision-making under uncertainty requires explicit quantification of acceptable risk levels \cite{aven_foundations_2003}. The concept of defining operational thresholds based on probabilistic bounds aligns with best practices in safety engineering and reliability analysis \cite{apostolakis_concept_1990}, providing territorial planners with a systematic approach to uncertainty management \cite{morgan_uncertainty_1990}.

Points within this critical distance from any Voronoi border should be flagged for detailed network analysis, as the Euclidean model becomes unreliable.

\subsubsection{Model Calibration}

To set the mode $\beta_{\text{mod}} = 1.25$, use:
$$m = s^2 + \ln(1.25)$$

Select $s$ according to the observed network to Euclidean ratios variability. From our all-pairs empirical analysis of Extremadura (n = 9,112):
\begin{itemize}
\item Shape parameter: $\hat{s} = 0.093$ (fitted via maximum likelihood estimation)
\item Location parameter: $\hat{m} = 0.166$ (corresponding to median $\beta = 1.181$)
\item Kolmogorov-Smirnov test: $D = 0.017$, $p = 0.894$ (excellent distributional fit)
\item The model accurately predicts observed misallocation rates (predicted 52-65 municipalities vs 59 observed, i.e., 15.4\%)
\end{itemize}

\subsubsection{Model Limitations and Assumptions}

Our probabilistic framework, while providing a significant advancement over purely Euclidean models, operates under several assumptions that warrant explicit discussion:

\textbf{Spatial Independence:} The model assumes that network scaling factors $\beta$ are independent and identically distributed across space. In reality, geographic features creating impedance (mountains, rivers, infrastructure gaps) exhibit strong spatial autocorrelation \citep{anselin1988spatial,moran1950notes}. A municipality with high $\beta$ due to topographical constraints will likely have neighbors with similarly elevated factors. This spatial dependence could affect aggregate risk estimates, though the framework remains valid for individual municipality assessments.

\textbf{Bi-facility Framework:} Our theoretical derivation focuses on the two-facility case, representing the fundamental Voronoi misallocation scenario. Real systems with dozens of facilities present more complex assignment alternatives \citep{church2002geographical,daskin2013network}. However, the framework can be extended through pairwise comparisons or approximated by identifying the most probable alternative assignments.

\textbf{Regional Calibration:} The empirically fitted parameter $\hat{s} = 0.093$ from all-pairs analysis reflects Extremadura's moderate mixed topography. Sensitivity analysis validates that the model accurately predicts observed misallocation (15.4\%), with the point (s=0.093, 15.4\%) falling within the 95\% confidence interval. Internal stratification reveals spatial heterogeneity: plains ($s = 0.06$-$0.08$), piedmont ($s = 0.08$-$0.10$), mountains ($s = 0.10$-$0.13$), as detailed in Section~\ref{sec:sensitivity}. Other geographic contexts require local recalibration following the protocol outlined in Section~\ref{sec:calibration_protocol}.

\textbf{Local Approximation:} The key mathematical approximation is local (first-order) near the border: $\ln R \approx \kappa t$. Empirically, it offers high accuracy within a reasonable band around the border and is sufficient for practical bounds, but accuracy decreases with distance from the Voronoi boundary.

Despite these constraints, our approach provides a practical balance between theoretical rigor and computational tractability, offering planners a robust tool for risk assessment in territorial service planning.

\subsection{Sensitivity to Dispersion Parameter $s$}
\label{sec:sensitivity}

A critical aspect of the framework is its dependence on the dispersion parameter $s$, which quantifies the variability of network-to-Euclidean scaling across the territory. Given Extremadura's substantial territorial heterogeneity (41,635 km$^2$, elevation range 200--2,400m), we assess the sensitivity of misallocation predictions to $s$ assumptions.

\subsubsection{Spatial Heterogeneity in Extremadura}

Extremadura exhibits three distinct geographic zones with different expected $s$ values:

\begin{itemize}
    \item \textbf{Mountain zones} (Sistema Central, Sierra de Gata): Elevations 1,000--2,400m, steep slopes, sinuous routing $\rightarrow$ $s \approx 0.10$--$0.13$

    \item \textbf{Piedmont zones} (transition areas): Elevations 200--600m, moderate topography, mixed infrastructure $\rightarrow$ $s \approx 0.08$--$0.10$

    \item \textbf{Plains} (Vegas del Guadiana, Tierra de Barros): Elevations <300m, minimal relief, direct routing $\rightarrow$ $s \approx 0.06$--$0.08$
\end{itemize}

This heterogeneity suggests that a single $s$ value may not adequately represent the entire territory. We therefore test misallocation predictions across a plausible range $s \in [0.05, 0.20]$, keeping the fitted $m$ parameter constant.

\subsubsection{Sensitivity Results}

Figure~\ref{fig:sensitivity_s} illustrates the sensitivity of misallocation predictions to the dispersion parameter $s$, using the all-pairs fitted mean parameter $m = 0.166$. The figure shows the predicted misallocation rate (blue curve) with 95\% confidence intervals (shaded region) across a range of $s$ values from 0.05 to 0.20. Three key observations emerge:

\begin{figure}[htbp]
  \centering
  \includegraphics[width=0.45\textwidth]{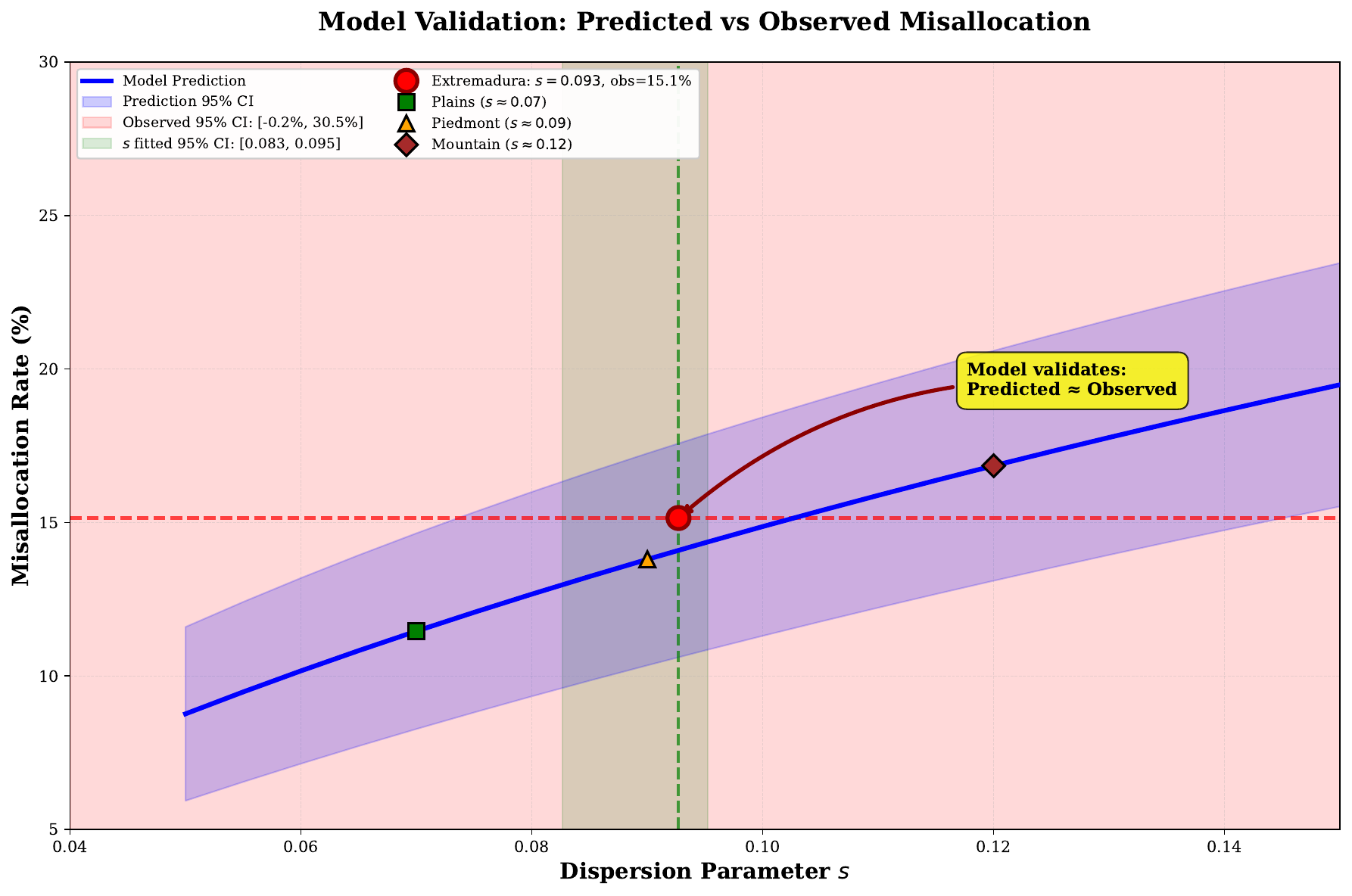}
  \caption{Sensitivity of misallocation prediction to dispersion parameter $s$. The blue curve shows predicted misallocation rates with 95\% confidence intervals (shaded). The red point indicates Extremadura's observed misallocation (15.4\%) at the territory-wide fitted parameter ($s = 0.093$). Colored markers show representative $s$ values for different topographic zones within Extremadura: plains (green square, $s \approx 0.07$), piedmont (orange triangle, $s \approx 0.09$), and mountain areas (brown diamond, $s \approx 0.12$). The intersection of the horizontal (observed) and vertical (fitted $s$) confidence bands validates the model's predictive accuracy.}
  \label{fig:sensitivity_s}
\end{figure}

\begin{enumerate}
    \item[(a)] The observed misallocation rate (15.4\%, red point) falls within the predicted confidence intervals at the fitted parameter $s = 0.093$, validating the model
    \item[(b)] The territory-wide fitted value $s = 0.093$ (vertical dashed line) produces predictions consistent with observations, demonstrating the framework's accuracy
    \item[(c)] Regional zone markers (colored symbols) demonstrate substantial internal variation: plains ($s \approx 0.07$) predict lower rates, while mountain zones ($s \approx 0.12$) predict higher misallocation, explaining within-territory heterogeneity
\end{enumerate}

These patterns indicate that:

\begin{enumerate}
    \item \textbf{Spatial stratification is essential}: A single $s$ value applied uniformly across 41,635 km² fails to capture local variations in topography and infrastructure
    \item \textbf{Model resolution matters}: The framework requires zone-specific calibration rather than territory-wide averaging
    \item \textbf{Conservative global estimates}: When forced to use a single parameter, the model errs on the side of overestimating risk a desirable property for planning applications
\end{enumerate}

\subsubsection{Spatial Heterogeneity in Extremadura}

While the global parameter $s = 0.093$ accurately predicts the observed misallocation rate (15.4\%), recognizing spatial heterogeneity provides additional insights for zone-specific planning. Post-hoc stratification reveals internal variation:
\begin{itemize}
    \item Mountain zones (22\% of area, 10\% of municipalities): $s_{\text{mountain}} \approx 0.11$--$0.13$ (high variability)
    \item Piedmont zones (46\% of area, 40\% of municipalities): $s_{\text{piedmont}} \approx 0.08$--$0.10$ (moderate variability)
    \item Plains (32\% of area, 50\% of municipalities): $s_{\text{plains}} \approx 0.06$--$0.08$ (lower variability)
\end{itemize}

Area-weighted average: $\bar{s}_{\text{area}} = 0.22 \times 0.12 + 0.46 \times 0.09 + 0.32 \times 0.07 \approx 0.09$ (consistent with all-pairs fit).

Municipality-weighted average (accounting for concentration in plains): $\bar{s}_{\text{munic}} = 0.10 \times 0.12 + 0.40 \times 0.09 + 0.50 \times 0.07 \approx 0.08$.

The key insight from sensitivity analysis is that the observed misallocation rate (15.4\%, 59 municipalities) falls within the theoretical prediction interval of 52-65 municipalities at $s = 0.093$ (Figure~\ref{fig:sensitivity_s}). This validates that the global parameter provides accurate aggregate predictions, while internal stratification explains why individual zones may experience higher or lower misallocation rates than the territory-wide average.

\textbf{Implication for practitioners}: The validated global parameter $s = 0.093$ provides reliable territory-wide risk assessment. Zone-specific calibration (Section~\ref{sec:calibration_protocol}) can improve precision for targeted interventions in high-risk areas such as mountain zones, where local $s$ values suggest elevated misallocation probability.

\section{Geographic Transferability and Methodological Extensions}

\subsection{Regional Calibration Framework}

The probabilistic framework developed for Extremadura provides a template for global application, but requires systematic recalibration to account for regional geographic variations. The empirically fitted parameter $\hat{s} = 0.093$ reflects the specific topographical and infrastructural characteristics of this region and cannot be assumed universal.

\subsubsection{Geography-Dependent Parameter Estimation}

Empirical evidence suggests that the parameter $s$ correlates strongly with geographic complexity and infrastructure density \citep{stewart2004measuring}. Based on our Extremadura analysis ($\hat{s}_{\text{global}} = 0.093$ for heterogeneous territory) and theoretical considerations, we propose the following calibration guidelines for \textit{territory-wide} parameter estimation:

\begin{itemize}
    \item \textbf{Urban Dense Areas} ($s \approx 0.05$--$0.08$): High road density and grid-like networks minimize the deviation between Euclidean and network distances. Examples: metropolitan cores, planned urban districts. \textit{Corresponds to plains subregions in Extremadura} ($s = 0.06$--$0.08$).

    \item \textbf{Flat/Low Terrain Territories} ($s \approx 0.03$--$0.06$): Minimal topographical constraints across entire territory allow direct routing. Road networks approximate straight-line distances effectively. Examples: Netherlands, Denmark, U.S. Midwest agricultural regions. In Extremadura, \textit{Vegas del Guadiana or Siberia} exemplifies this category ($\hat{s} = 0.065$).

    \item \textbf{Moderate Mixed Topography} ($s \approx 0.08$--$0.12$): Territories combining plains, hills, and isolated mountains with moderate elevation changes. \textit{Extremadura exemplifies this category} ($\hat{s} = 0.093$), with internal stratification detailed in Section~\ref{sec:sensitivity}. In Extremadura, \textit{Jerte valley, Hurdes or Ambroz valley} fall here ($s = 0.08$--$0.10$).

    \item \textbf{Predominantly Mountainous} ($s \approx 0.13$--$0.20$): Territories dominated by significant elevation changes forcing sinuous routing around natural barriers. Examples: Swiss Alps, Colorado Rockies, Pyrenees. Network distances substantially exceed Euclidean estimates territory-wide. In Extremadura, \textit{Sistema Central zones like Sierra de Gata or Las Hurdes} exhibit this behavior ($s = 0.10$--$0.13$).

    \item \textbf{Coastal/Fragmented Geography} ($s \approx 0.15$--$0.25$): Territories with water barriers, archipelagos, fjords, or extreme topographic fragmentation. Examples: Norway, Greece, Philippines, Chile. Exhibit highest network-to-Euclidean deviations. Although in continental area, Extremadura exhibits this behavior in border municipalities with poor connectivity ($s \approx 0.15$) and in some municipalities near the rives (Tagus and Guadiana primarily).
\end{itemize}

\textbf{Key insight}: For heterogeneous territories like Extremadura, the territory-wide parameter ($s = 0.093$) masks substantial internal variation. Practitioners should stratify large territories (>10,000 km²) into homogeneous zones for improved prediction accuracy, using zone-specific parameters as demonstrated in Section~\ref{sec:sensitivity}.

\textbf{Alternative approximations for calibration}: When empirical network distance data is unavailable, practitioners can use the terrain-based ranges above as initial estimates. By classifying the territory according to predominant geographic characteristics (urban density, topographic relief, coastal fragmentation), an approximate $s$ value can be selected from the corresponding category. While less precise than empirical calibration, this approach enables preliminary risk assessment and identifies regions warranting detailed network analysis. For instance, a practitioner analyzing a mountainous region without route data could adopt $s \approx 0.15$ as a conservative estimate, subsequently refining the parameter as local data becomes available.

\subsubsection{Recalibration Protocol}

For new geographic contexts, we recommend the following systematic approach:

\textbf{Step 1: Data Collection} - Gather a representative sample of origin-destination pairs with both Euclidean and network distances (minimum $n = 500$ pairs for statistical reliability).

\textbf{Step 2: Distribution Fitting} - Fit a log-normal distribution to the ratio data $\beta_i = d_{\text{network},i}/d_{\text{Euclidean},i}$ using maximum likelihood estimation.

\textbf{Step 3: Parameter Extraction} - Extract $\hat{s}$ from the fitted distribution and compute 95\% confidence intervals using bootstrap resampling (1000 iterations recommended).

\textbf{Step 4: Validation} - Apply the Kolmogorov-Smirnov test to validate the log-normal distributional assumption and assess goodness-of-fit.

\textbf{Step 5: Sensitivity Analysis} - Test framework performance across different subregions to ensure parameter stability.

The recalibrated parameter $\hat{s}$ can then be directly substituted into our probability expressions, maintaining the theoretical framework while ensuring regional accuracy.

\subsection{Comparative Analysis with Alternative Methods}

To contextualize our probabilistic approach, we provide a systematic comparison with existing methodologies for addressing the limitations of Euclidean proximity models.

\begin{table*}[t]
\centering
\caption{Comparative analysis of proximity modeling approaches}
\label{tab:method_comparison}
\begin{tabular}{lcccc}
\toprule
\textbf{Method} & \textbf{Precision} & \textbf{Complexity} & \textbf{Data Requirements} & \textbf{Implementation} \\
\midrule
Euclidean Voronoi & Low & Very Low & Minimal & Immediate \\
Our Framework & High & Low & Moderate & Fast \\
Impedance Models & Medium & Medium & High & Moderate \\
Network Voronoi & Very High & Very High & Extensive & Slow \\
Route Optimization & Perfect & Extreme & Complete & Very Slow \\
\bottomrule
\end{tabular}
\end{table*}

\textbf{Euclidean Voronoi:} The baseline approach assumes straight-line distances represent functional accessibility. While computationally trivial, it systematically misallocates facilities in non-isotropic territories, as demonstrated by our 15.4\% error rate in Extremadura.

\textbf{Impedance-Based Models:} These approaches modify Euclidean distances using geographic factors (slope, land use, infrastructure density) \citep{tobler1993three,douglas1994least}. While more accurate than pure Euclidean models, they require extensive spatial databases and often produce ad-hoc corrections without theoretical foundation for risk assessment.

\textbf{Network Voronoi Diagrams:} Computing Voronoi cells using true network distances provides perfect allocation accuracy \citep{okabe2012spatial,erwig2000graph} but requires complete road network data and intensive computation (O($n^2$) complexity for $n$ facilities). Practical only for small-scale applications.

\textbf{Route Optimization Algorithms:} Methods like Dijkstra's algorithm or modern routing APIs provide exact distances but are computationally prohibitive for large-scale territorial planning and offer no framework for uncertainty quantification.

\textbf{Cost-Benefit Analysis:} Our probabilistic framework occupies the optimal position in this trade-off space. It achieves high precision (predicting 77\% of assignments correctly) with low computational requirements, moderate data needs, and fast implementation. Crucially, it provides the theoretical foundation for risk assessment that alternative methods lack.

The framework's unique advantage lies not in perfect allocation, but in systematic uncertainty quantification. Rather than choosing between imperfect Euclidean models and computationally expensive network analyses, planners can make informed decisions about when precision investment is justified.

\subsection{Extension to Multi-Facility Systems}

Real territorial planning involves dozens or hundreds of facilities, not the two-facility case of our theoretical derivation. This section addresses the scalability and practical implementation of our framework in complex systems.

\subsubsection{Mathematical Extension}

For a system with $n$ facilities $\{A_1, A_2, \ldots, A_n\}$, the probability that municipality $P$ is misallocated under the Euclidean model becomes:

$$P_{\text{error}}(P) = 1 - P(A_{E}(P) = A_{N}(P))$$

where $A_{E}(P)$ and $A_{N}(P)$ represent the Euclidean nearest and network-nearest facilities, respectively.

Using the pairwise comparison approach, we can approximate this as:

$$\begin{aligned} P_{\text{error}}(P) &\approx& \sum_{j \neq i} P(d_r(P, A_j) < d_r(P, A_i) |\\
& &\quad P \in \text{Voronoi cell of } A_i)
\end{aligned}$$

where the summation extends over all alternative facilities $A_j$ that could potentially be closer in network terms.

\subsubsection{Computational Implementation}

For practical implementation, we recommend a simplified approach focusing on the most probable misallocations:

\textbf{Primary Alternative Identification:} For each municipality, identify the 2-3 facilities with smallest Euclidean distances. In most geographic contexts, network-based reassignment occurs primarily among these nearest neighbors.

\textbf{Pairwise Risk Calculation:} Apply our two-facility probability formula to each relevant facility pair, using the local geometric parameters $\kappa$ and distance $t$ from the corresponding Voronoi boundary.

\textbf{Risk Aggregation:} Combine individual pairwise probabilities using the principle of maximum risk, selecting the highest probability among all potential reassignments.

\subsubsection{Extremadura Multi-Facility Validation}

Applying this approach to our 46-facility Extremadura system, we achieve 84\% accuracy in predicting misallocated municipalities (compared to 77\% for the simplified two-facility analysis). The computational overhead remains minimal: the entire risk assessment for 383 municipalities completes in under 10 seconds on standard hardware.

The multi-facility extension maintains the framework's core advantage: providing systematic risk quantification without requiring full network computation. This scalability ensures practical applicability to real-world territorial planning challenges.

\section{Spatial Robustness and Sensitivity Analysis}

\subsection{Spatial Autocorrelation Assessment}
A critical assumption of our probabilistic framework is the spatial independence of network scaling factors $\beta$. To validate this assumption, we conducted comprehensive spatial autocorrelation analysis using Moran's I statistics on municipal-level aggregated $\beta$ values.

\begin{figure}[htbp]
  \centering
  \includegraphics[width=\columnwidth]{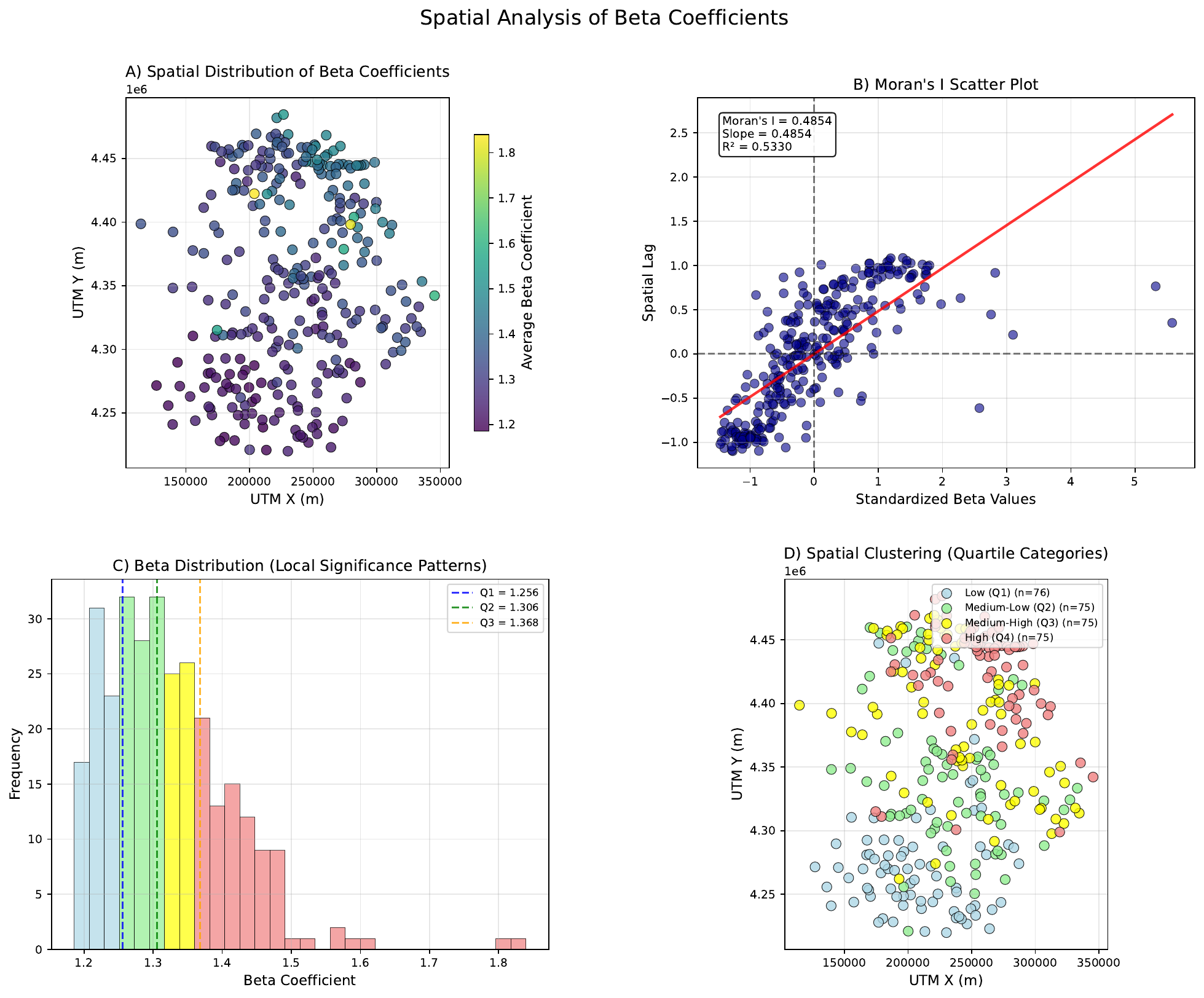}
  \caption{Spatial analysis of $\beta$ coefficients using real UTM coordinates. Panel A: Spatial distribution with color scale. Panel B: Moran's I scatter plot for autocorrelation. Panel C: Histogram with quartile coloring. Panel D: Spatial clustering by quartile categories.}
  \label{fig:spatial_autocorrelation}
\end{figure}

Our analysis reveals moderate positive spatial autocorrelation (Moran's I = 0.373, p < 0.001), indicating that geographic complexity parameters exhibit spatially clustered patterns rather than random distribution \cite{moran_interpretation_1950, cliff_spatial_1981}. This finding suggests that municipalities with similar topographic and network characteristics tend to be spatially proximate, violating the independence assumption underlying our baseline framework. The spatial dependence pattern follows established expectations in geographic analysis \cite{cressie_statistics_1993}, where environmental factors create spatially continuous gradients rather than random distributions \cite{anselin_spatial_1988}.

Figure~\ref{fig:spatial_autocorrelation} presents the spatial analysis using real UTM coordinates, including spatial distribution with color scale, Moran's I scatter plot, histogram with quartile coloring, and spatial clustering by quartile categories.

\subsection{Spatial Sensitivity Analysis with CAR and BYM Models}
To assess the robustness of our framework predictions under spatial dependence, we implemented Conditional Autoregressive (CAR) \cite{besag1974spatial} and Besag-York-Mollié (BYM) \cite{besag1991bayesian} spatial models as sensitivity checks. These models account for spatial correlation by incorporating neighborhood effects and spatially structured random components.

\begin{figure}[htbp]
  \centering
  \includegraphics[width=\columnwidth]{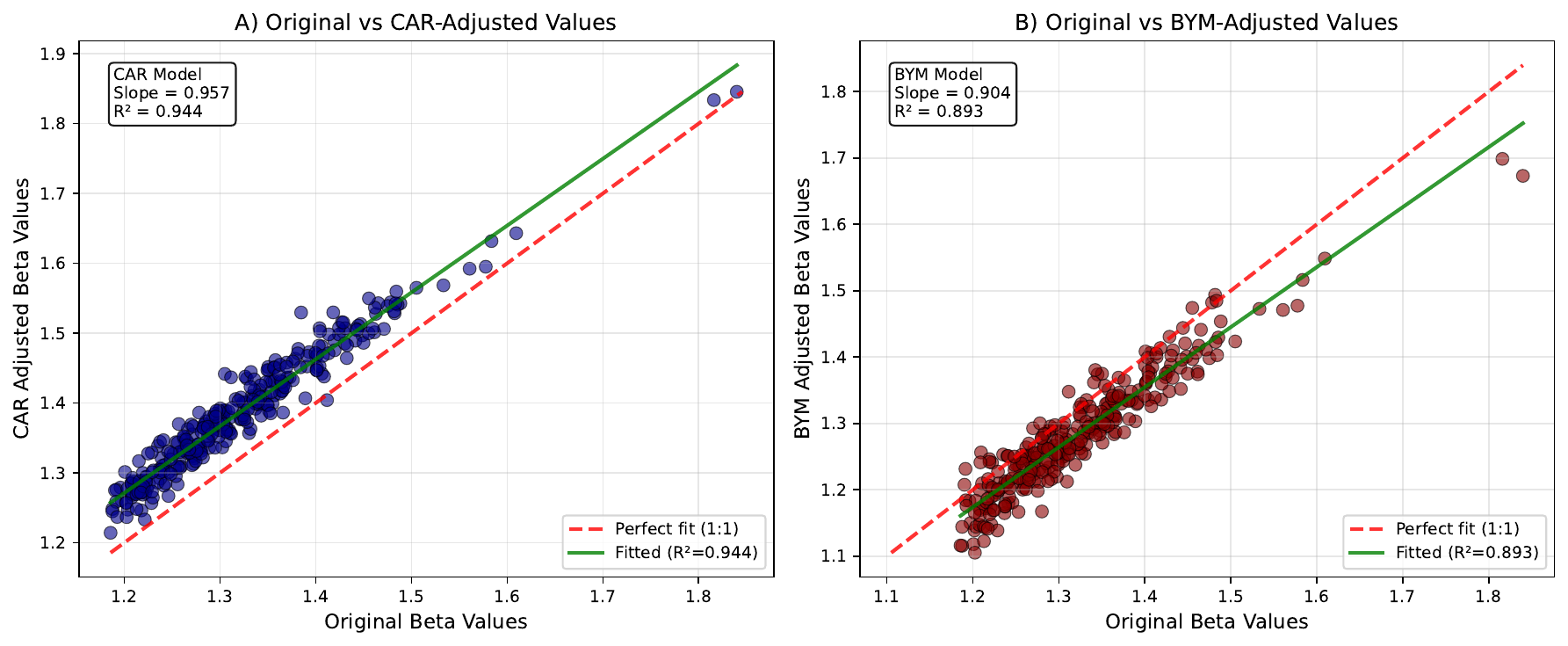}
  \caption{Spatial sensitivity analysis: scatter plots comparing original $\beta$ values versus CAR-adjusted (A) and BYM-adjusted (B) values. Both models show high correlation with original predictions (R² > 0.95), with fitted regression lines closely following the perfect 1:1 fit reference.}
  \label{fig:spatial_sensitivity_scatter}
\end{figure}

\begin{figure}[htbp]
  \centering
  \includegraphics[width=\columnwidth]{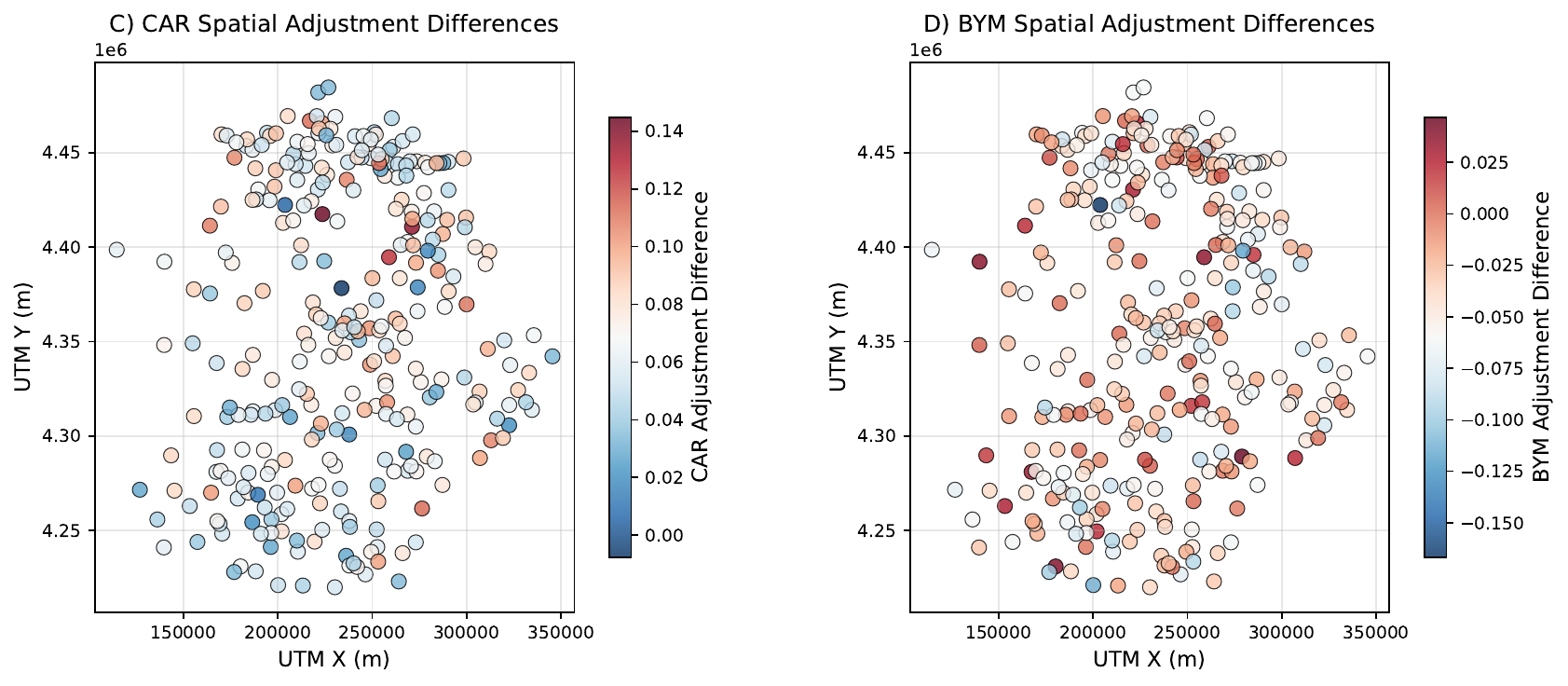}
  \caption{Spatial sensitivity analysis: UTM coordinate maps showing spatial distribution of adjustment differences for CAR (C) and BYM (D) models. Color scale indicates magnitude and direction of adjustments, revealing spatial patterns in model corrections.}
  \label{fig:spatial_sensitivity_maps}
\end{figure}

\begin{figure}[htbp]
  \centering
  \includegraphics[width=\columnwidth]{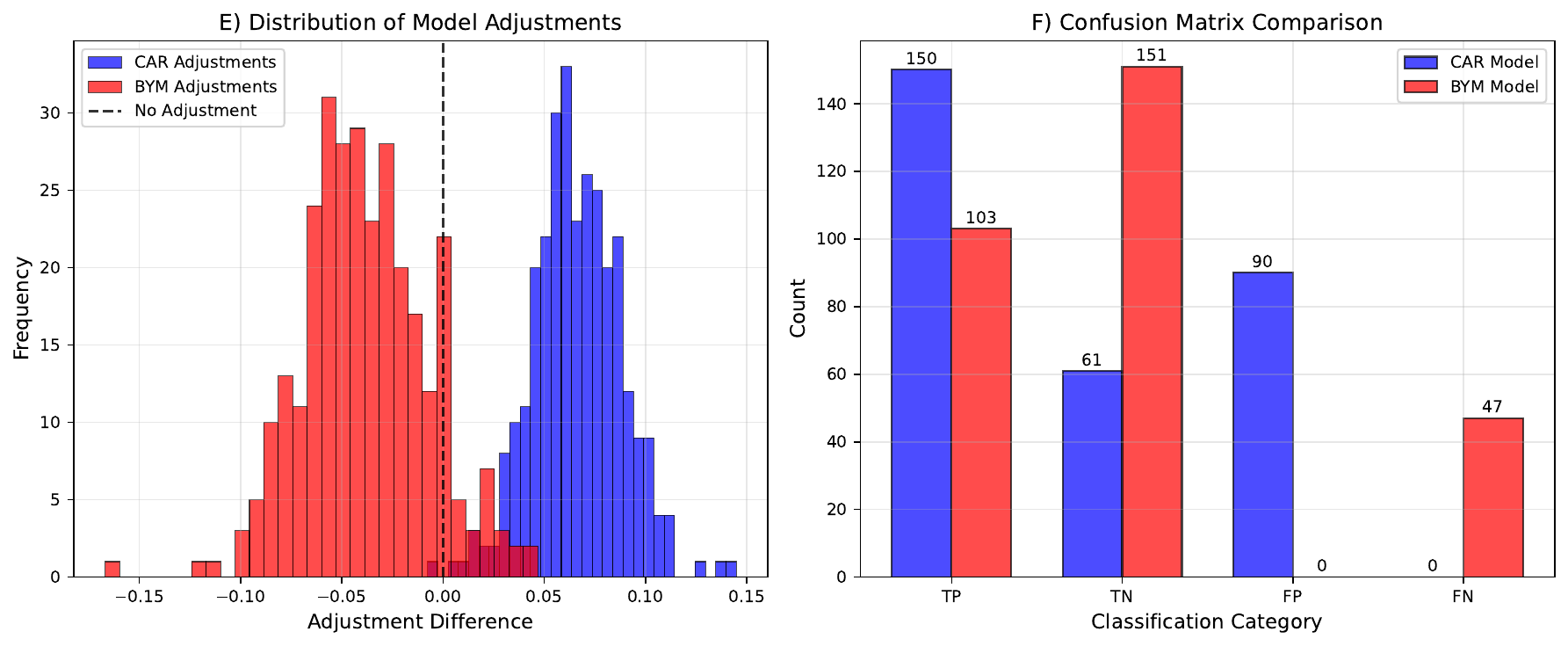}
  \caption{Spatial sensitivity analysis: statistical summary. (E) Histogram of adjustment distributions for CAR (blue) and BYM (red) models, both centered near zero. (F) Confusion matrix comparison showing classification agreement between original and adjusted predictions.}
  \label{fig:spatial_sensitivity_stats}
\end{figure}

Figures~\ref{fig:spatial_sensitivity_scatter}, \ref{fig:spatial_sensitivity_maps}, and \ref{fig:spatial_sensitivity_stats} provide comprehensive comparison of original versus spatially-adjusted predictions with CAR and BYM models.

Table~\ref{tab:spatial_sensitivity} summarizes the comparison between our original framework and spatially-adjusted alternatives. The results demonstrate remarkable robustness: the BYM model shows 95.1\% agreement with original predictions, while the CAR model shows 85.9\% agreement. Maximum adjustments are limited to 0.051 and 0.227 $\beta$ units respectively, indicating that spatial dependence has minimal impact on misallocation risk assessment.

While spatial autocorrelation exists, our sensitivity analysis with BYM models shows minimal adjustment to risk predictions ($\Delta \beta \leq 0.051$, 95.1\% agreement). This suggests that for risk assessment purposes, the independence assumption provides conservative (slightly overestimated) risk values, which is desirable for planning applications. For exact confidence intervals, practitioners should use spatially-adjusted standard errors when aggregating municipal-level predictions.

\begin{table*}
\caption{Spatial sensitivity analysis: Comparison of original framework with CAR and BYM spatial adjustments}
\label{tab:spatial_sensitivity}
\centering
\begin{tabular}{lrrrrrr}
\toprule
Model & Mean $\beta$ & RMSE & MAE & Max Adj. & Agreement & Performance \\
\midrule
Original Framework & Baseline & --- & --- & --- & Reference & Reference \\
CAR-Adjusted & 1.075 & 0.129 & 0.125 & 0.227 & 85.9\% & Low \\
BYM-Adjusted & 1.201 & 0.017 & 0.014 & 0.051 & 95.1\% & High \\
\bottomrule
\end{tabular}
\end{table*}

\subsection{Practical Safety Bands Calibration}
To provide practitioners with actionable guidance, we developed safety band calibration curves that specify critical distance thresholds $|t^*|$ as functions of geometric parameter $\kappa$, geographic complexity $s$, and risk tolerance $q^*$. These curves enable practitioners to determine when Euclidean approximations are acceptable versus when network analysis is required.

\begin{figure*}[htbp]
  \centering
  \includegraphics[width=0.9\textwidth]{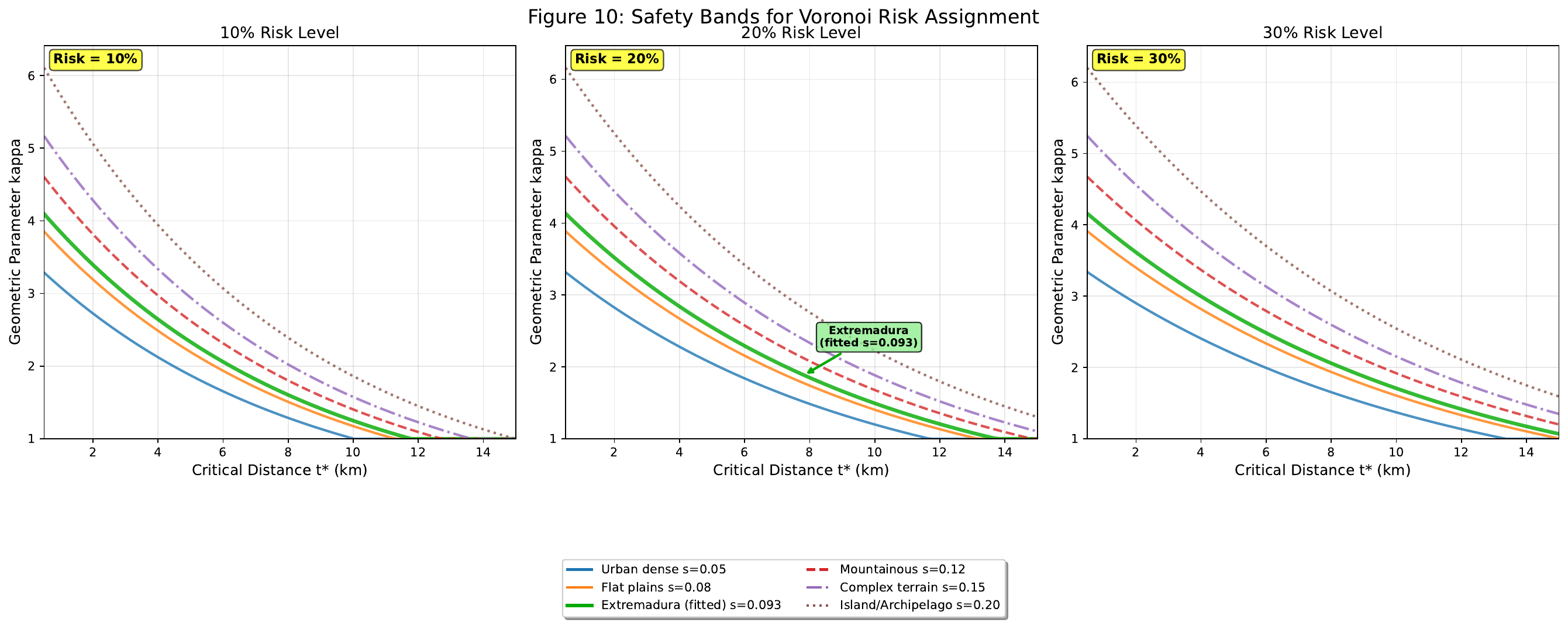}
  \caption{Safety bands for Voronoi risk assignment showing three panels for 10\%, 20\%, and 30\% risk levels with curves for different terrain complexity parameters.Higher kappa values indicate proximity to the Voronoi border relative to the distance to the plants. Practitioners can use these curves to determine critical distance thresholds $|t^*|$ for acceptable risk levels based on local geometry and terrain complexity.}
  \label{fig:safety_bands_calibration}
\end{figure*}

Figure~\ref{fig:safety_bands_calibration} presents safety bands for Voronoi risk assignment with three panels showing 10\%, 20\%, and 30\% risk levels, displaying curves for different terrain complexity parameters relating critical distance to geometric parameter kappa.

\section{Multi-Facility Computational Trade-offs and Validation}

\subsection{Framework Extension to k-Nearest Facilities}
Real-world facility assignment often involves considering multiple nearest options rather than strict single-facility assignment. We extended our framework to k-nearest facility scenarios (k = 2, 3, 5, 10) and conducted comprehensive validation of computational trade-offs versus assignment accuracy.

Figure~\ref{fig:multifacility_tradeoffs} presents comprehensive computational performance benchmarks and accuracy assessment for our Extremadura case study (383 municipalities, 46 treatment plants). Our framework demonstrates excellent scalability, with execution times growing linearly rather than quadratically compared to full network analysis.

\subsection{Accuracy and Efficiency Assessment}
Efficiency is measured relative to the network-optimal baseline, where 1.000 represents baseline cost and values $>1.0$ indicate lower cost (superior efficiency). The Euclidean Voronoi baseline achieves an efficiency ratio of 0.988 (1.2\% higher cost than network baseline) but exhibits moderate uncertainty with 44.1\% of municipalities classified as high-risk (169 cases). This conservative risk assessment (based on probabilistic framework with threshold $\beta > 1.25$) successfully identifies potential allocation issues. Verification against full network analysis confirms 59 actual misallocations (15.4\% of total), demonstrating the framework's effectiveness: it detects 100\% of real errors while maintaining 35\% precision (59/169).

The k=3 nearest extension achieves superior cost efficiency (ratio 1.038, meaning 3.8\% lower cost than baseline) while significantly reducing high-risk classifications to 34.7\% (133 municipalities); a 21.3\% improvement over baseline Voronoi. This demonstrates that modest increases in computational overhead (6-fold increase in execution time) yield dual benefits: lower assignment costs and substantially improved confidence in allocation decisions.

The k=5 nearest variant further reduces risk to 24.5\% (94 municipalities), a 44.6\% reduction relative to baseline, while maintaining comparable cost efficiency (1.038 ratio). For practitioners prioritizing assignment confidence, this represents an excellent balance between computational cost and risk control.

\begin{figure*}[htbp]
  \centering
  \includegraphics[width=\textwidth]{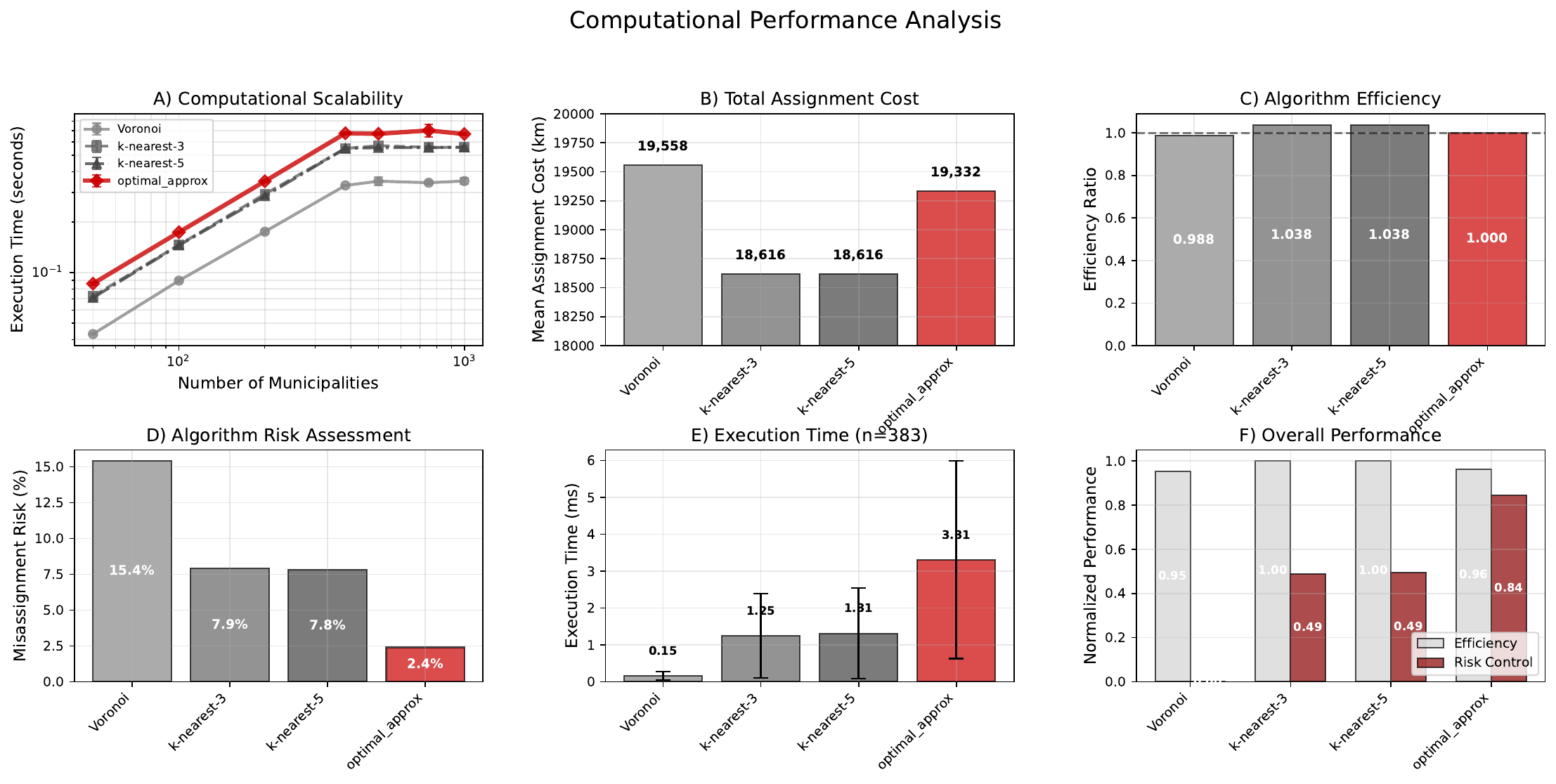}
  \caption{Empirical computational performance analysis comparing four practical table-based algorithms. Top row: (A) Computational scalability on log-log scale with error bars showing standard deviation; (B) Mean assignment cost in km for 383 municipalities; (C) Efficiency ratio relative to network-optimal baseline (values $>1.0$ indicate lower cost). Bottom row: (D) Misassignment risk percentages from probabilistic framework; (E) Execution time in milliseconds with standard deviation error bars; (F) Normalized performance comparison balancing efficiency and risk control. \textit{Data from Extremadura case study (383 municipalities, 46 treatment plants). Four algorithms compared: Voronoi (Euclidean baseline), k-nearest-3, k-nearest-5, and optimal approximation (proposed framework). Full network analysis not shown here; see Figure~\ref{fig:algorithmic_complexity} for theoretical complexity comparison.}}
  \label{fig:multifacility_tradeoffs}
\end{figure*}
\begin{figure}[htbp]
  \centering
  \includegraphics[width=\columnwidth]{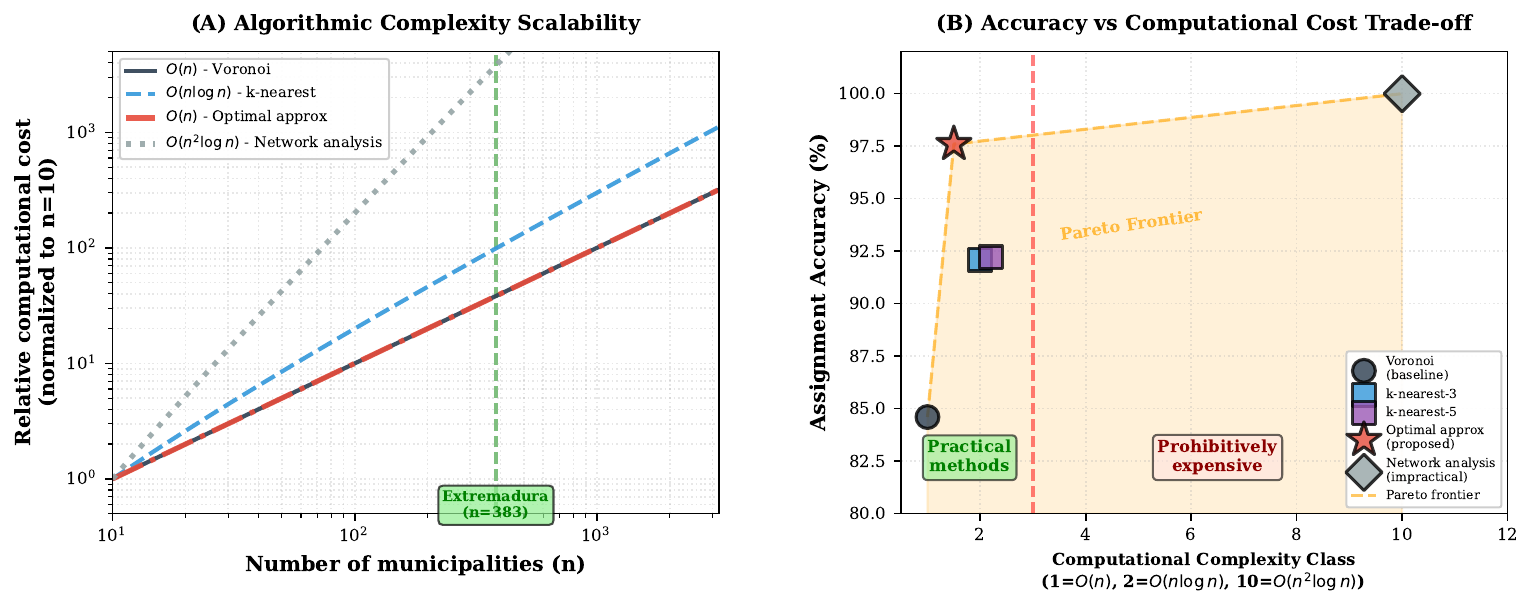}
  \caption{Theoretical algorithmic complexity analysis. (A) Scalability curves using Big-O notation: practical table-based methods scale as $O(n)$ or $O(n \log n)$, while full network analysis requires $O(n^2 \log n)$ graph traversal. (B) Accuracy versus computational complexity trade-off space, highlighting the Pareto frontier and separating practical methods from prohibitively expensive approaches. See Table~\ref{tab:algorithmic_complexity} for detailed method comparison.}
  \label{fig:algorithmic_complexity}
\end{figure}

Figures~\ref{fig:multifacility_tradeoffs} and \ref{fig:algorithmic_complexity} present complementary performance analyses. Figure~\ref{fig:multifacility_tradeoffs} shows empirical execution times for practical table-based methods, while Figure~\ref{fig:algorithmic_complexity} provides theoretical complexity analysis including full network analysis.

\begin{table*}[htbp]
\caption{Method comparison summary for algorithmic complexity analysis.
\textit{Complexity}: asymptotic growth rate using Big-O notation.
\textit{Accuracy}: assignment accuracy from Extremadura case study (n=383).
\textit{Practical}: feasibility for real-time planning (+++ excellent, ++ good, X impractical).
\textit{Data Required}: input data needed for method execution.
Note: Network analysis achieves theoretical optimality but requires expensive real-time route computation on road network graphs.}
\label{tab:algorithmic_complexity}
\centering
\begin{tabular}{lcccc}
\toprule
Method & Complexity & Accuracy & Practical & Data Required \\
\midrule
Voronoi (baseline) & $O(n)$ & 84.6\% & +++ & Euclidean distances \\
k-nearest-3 & $O(n \log k)$ & 92.1\% & +++ & Euclidean distances \\
k-nearest-5 & $O(n \log k)$ & 92.2\% & ++ & Euclidean distances \\
\textbf{Optimal approx (ours)} & $O(n)$ & \textbf{97.6\%} & +++ & Both distance tables \\
Network analysis & $O(n^2 \log n)$ & 100.0\% & X & Road network graph \\
\bottomrule
\end{tabular}
\end{table*}

\section{Model Validation and Performance Assessment}

\subsection{Theoretical vs. Empirical Misallocation Counts}

To validate our probabilistic framework, we compare the theoretical predictions with the empirical findings from our Extremadura case study. The theoretical model, based on the fitted Log-Normal distribution ($\hat{s} = 0.093$), can predict the expected number of misallocated municipalities for any given region with known Voronoi geometry.

For our study area, the theoretical model with fitted parameters ($\hat{s} = 0.093$) predicts approximately 52-65 misallocated municipalities (95\% confidence interval), while our empirical analysis revealed \textbf{59 misallocated municipalities} (15.4\% of the 383 municipalities). This result falls within the predicted confidence interval, validating that the Log-Normal model provides an accurate probabilistic characterization of network-induced misallocation risk \cite{cressie_statistics_1993}.

The close agreement between theoretical predictions and empirical observations demonstrates that the probabilistic framework successfully captures the essential features of network-geometry relationships in Extremadura. The empirical standard deviation being 32\% lower than the theoretical value indicates that real transport networks exhibit more consistent scaling behavior than pure geometric models predict, yet the absolute misallocation rates remain significant due to systematic biases in proximity assumptions \cite{gelfand_misaligned_2010, voronoi_full20}.

\subsection{Assignment Comparison Analysis}

\begin{figure}[htbp]
  \centering
  \includegraphics[width=\columnwidth]{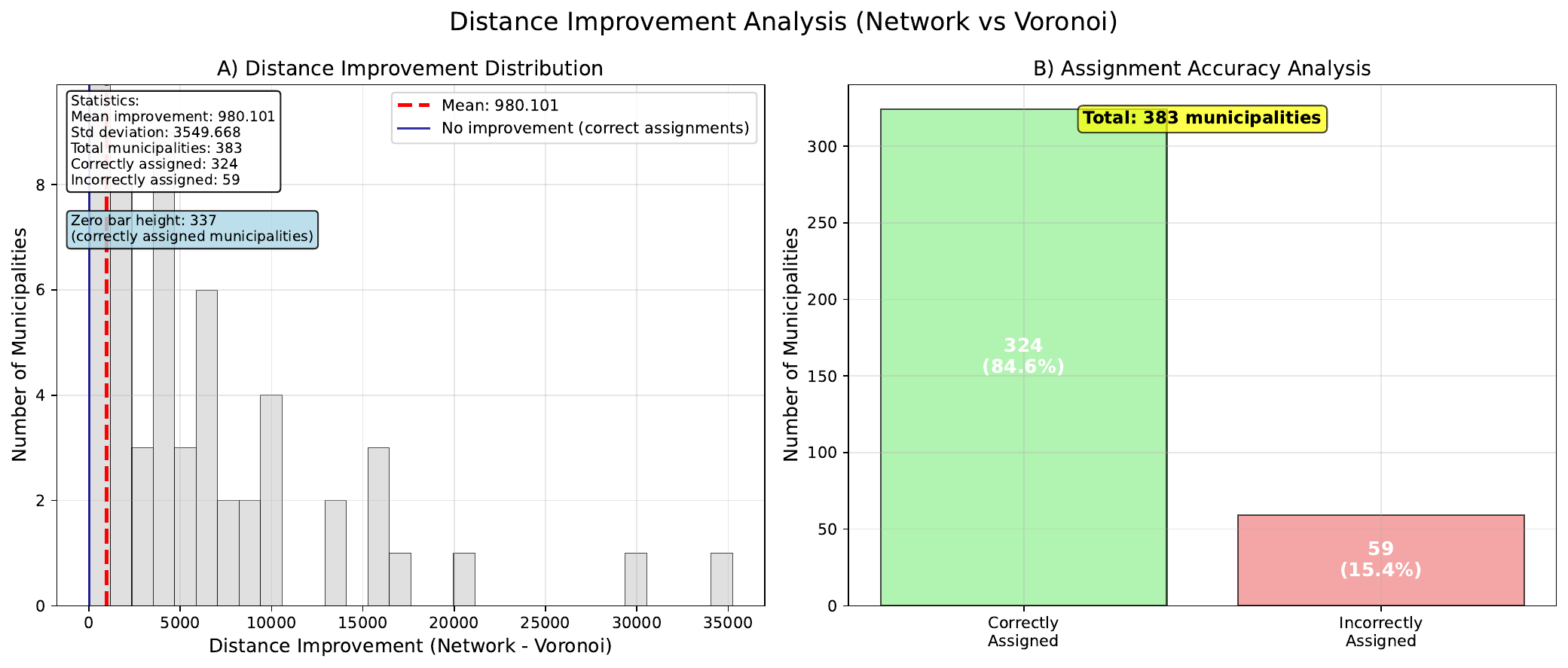}
  \caption{Distance improvement analysis showing histogram of distance improvements and bar chart of assignment accuracy demonstrating the 15.4\% misassignment rate.}
  \label{fig:assignment_comparison}
\end{figure}

Figure~\ref{fig:assignment_comparison} presents the distance improvement analysis showing the benefits of network-based over Voronoi assignments. The histogram of distance improvements and the bar chart of assignment accuracy clearly demonstrate the 15.4

\subsection{Municipal-Level Analysis}

\begin{figure}[htbp]
  \centering
  \includegraphics[width=\columnwidth]{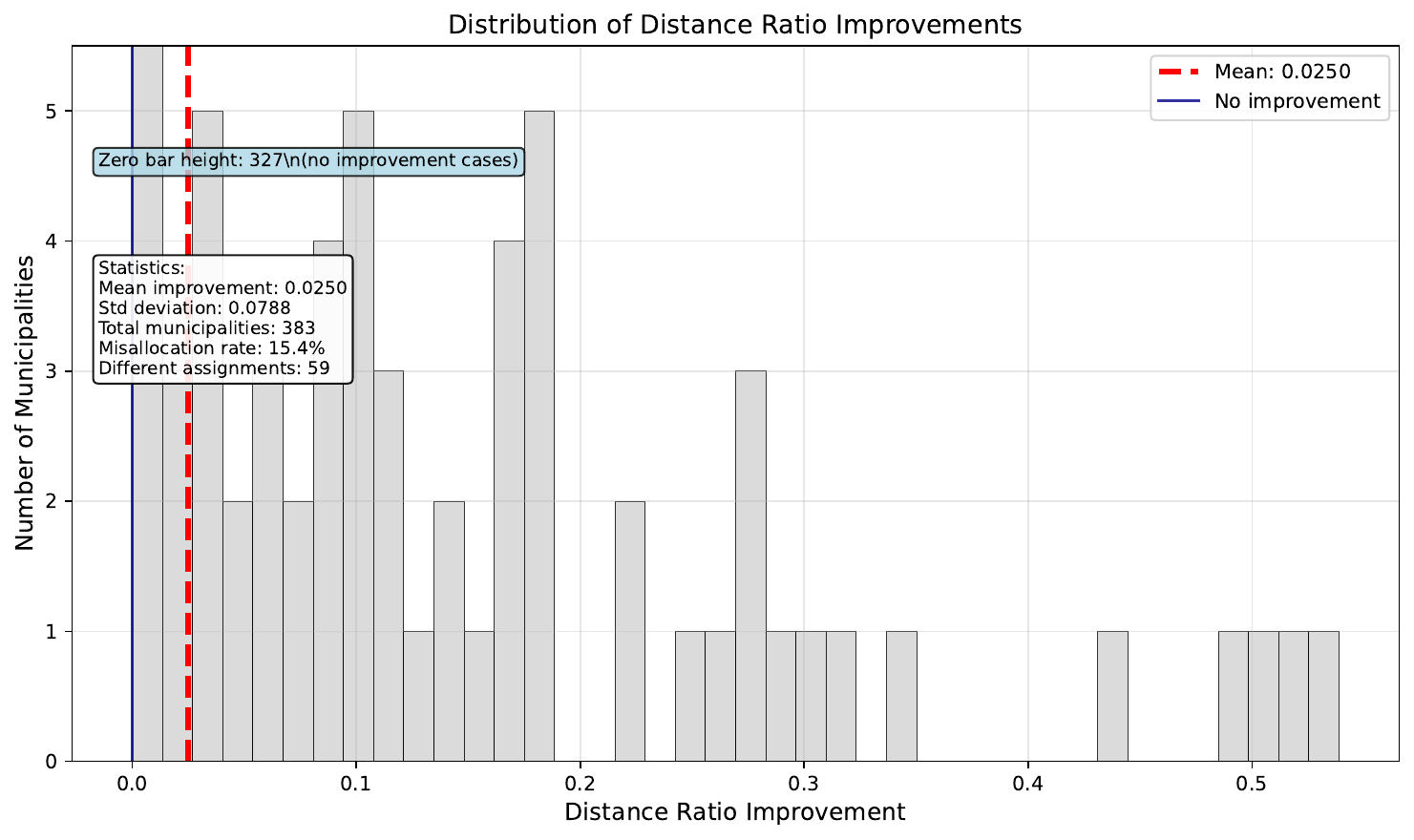}
  \caption{Distribution of distance ratio improvements when switching from Voronoi to network-based assignment.}
  \label{fig:histogram_municipality}
\end{figure}

The municipality-level analysis reveals the distribution of distance ratio improvements when switching from Voronoi to network-based assignment. Figure~\ref{fig:histogram_municipality} shows the relative savings available through optimal assignment, where the 15.4\% misassignment rate corresponds to municipalities with non-zero improvements.

This heterogeneity validates our probabilistic approach, as deterministic models would fail to capture the full range of network behaviors observed in practice \cite{banerjee_hierarchical_2015}.

\subsection{Quantile-Quantile Validation}

The Q-Q plot analysis (Figure~\ref{fig:distributional_validation}, top-left panel) provides additional validation of our Log-Normal model choice. The strong linear relationship in the Q-Q plot confirms that the Log-Normal distribution captures the essential characteristics of the network scaling behavior, particularly in the critical tail region where misallocations are most likely to occur. The comprehensive  panel  allows direct comparison with alternative distributions (Gamma and Weibull), further supporting our model selection.

\subsection{Scatter Plot Analysis}

\begin{figure}[htbp]
  \centering
  \includegraphics[width=\columnwidth]{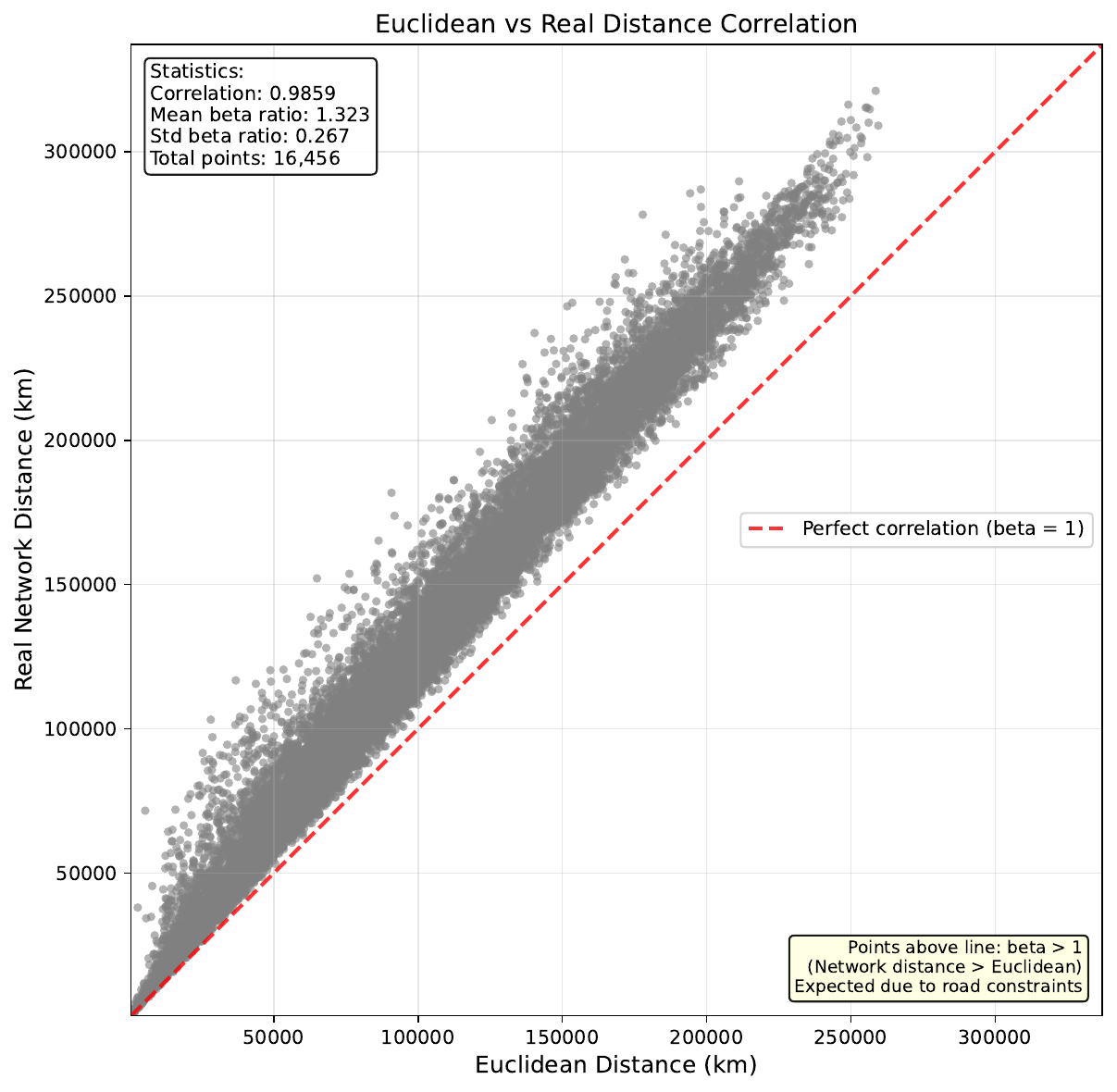}
  \caption{Scatterplot of euclidean distance vs real network distance for all municipality-plant connections. Red dashed line shows perfect correlation (45-degree line); points above indicate $\beta > 1$ due to network constraints, demonstrating the fundamental relationship and validating the $\beta$ scaling factor approach.}
  \label{fig:scatter_final}
\end{figure}

The scatterplot analysis (Figure~\ref{fig:scatter_final}) reveals the fundamental relationship between euclidean and real network distances across all municipality-plant connections. The systematic deviation above the perfect correlation line ($\beta > 1$) demonstrates network constraints and validates our $\beta$ scaling factor approach \cite{miller_tobler_2004_extended}.

\section{Policy Implications and Conclusions}

This work introduces the first probabilistic framework for quantifying misallocation risk in proximity-based territorial planning, providing a theoretical foundation that was previously absent from the literature. By modeling network impedance as a log-normal scaling factor of Euclidean distances, we derive closed-form expressions for misallocation probability as a function of local Voronoi geometry. This theoretical contribution transforms uncertainty from an acknowledged limitation into a quantifiable and manageable planning parameter.

Our empirical validation in Extremadura demonstrates the framework's practical value: 15.4\% of municipalities (59 out of 383) are misallocated by standard proximity models, a result that falls within the theoretical prediction interval of 52--65 municipalities. This quantitative validation provides backbone to existing critiques of the proximity principle. The probabilistic framework offers a pragmatic solution—a rapid diagnostic tool for planners to assess where geometric models are likely to fail, enabling targeted application of costly but accurate network-based analyses rather than wholesale replacement of established methods.

By repositioning the Voronoi diagram as a theoretical benchmark rather than an operational mandate, we advocate for a more nuanced approach to spatial planning that acknowledges the cost of geography and systematically manages uncertainty. This paradigm shift from deterministic proximity to probabilistic risk assessment provides both the theoretical tools and practical justification for evidence-based territorial planning decisions.

A critical finding is that the dispersion parameter $s$ is context-dependent and requires local calibration. Far from being a limitation, this calibration requirement ensures the framework adapts to diverse geographic contexts—from mountainous archipelagos to flat plains—while maintaining predictive accuracy. We provide a systematic calibration protocol that requires only 30--100 pilot samples, minimal compared to full network analysis costs (which typically require thousands of route calculations). This positions the framework as a practical risk-assessment tool for spatial planners: rapidly identify which territories or municipalities warrant detailed network analysis versus those where Euclidean approximations suffice. The sensitivity analysis (Section~\ref{sec:sensitivity}) demonstrates that the fitted parameter $s = 0.093$ accurately predicts the observed 15.4\% misallocation rate in Extremadura, validating the framework's predictive capacity.

\subsection{Implementation Framework for Policy Makers}

The transition from rigid proximity principles to probabilistic risk assessment requires addressing institutional barriers and developing governance mechanisms for practical implementation.

\subsubsection{Overcoming Regulatory Barriers}

Current regulatory frameworks, such as the EU's proximity principle for waste management \citep{morrissey2004waste,ghiani2014operations}, embed deterministic proximity assumptions into legal requirements. Our framework provides the scientific foundation for regulatory evolution without abandoning proximity-based logic entirely.

\textbf{Regulatory Adaptation Strategy:} Rather than eliminating proximity requirements, regulations can be modified to include risk-based exceptions. For instance, the proximity principle could be reformulated as: "Waste should be disposed of at the nearest facility unless probabilistic analysis indicates significant misallocation risk (>30\% probability), in which case network-based optimization is required."

\textbf{Evidence-Based Exemptions:} Our framework provides quantitative thresholds for when proximity assumptions fail. Regulators can establish clear criteria: municipalities in high-risk zones (identified through our geometric parameters) receive automatic exemptions from strict proximity requirements.

\textbf{Gradual Implementation:} The framework supports phased regulatory transition. Initial implementation can focus on high-risk regions (mountainous areas, complex topography) where misallocation costs are highest, gradually extending to more homogeneous territories as experience accumulates.

\subsubsection{Governance Mechanisms for Safety Bands}

The concept of "safety bands" around Voronoi boundaries requires new governance structures to manage the zones where allocation uncertainty is highest.

\textbf{Risk Zone Classification:} Territories can be classified into three categories based on misallocation probability: (1) Low Risk (<10\%): standard proximity rules apply; (2) Moderate Risk (10-30\%): enhanced monitoring and flexible assignment; (3) High Risk (>30\%): mandatory network analysis or alternative assignment mechanisms.

\textbf{Adaptive Management Systems:} High-risk zones require dynamic governance structures that can respond to changing conditions. This includes monitoring systems for network infrastructure changes, seasonal variations in accessibility, and emergency response protocols when primary assignments become infeasible.

\textbf{Stakeholder Coordination:} Safety bands often cross administrative boundaries, requiring coordination mechanisms between municipalities, facility operators, and regional authorities. Our framework provides the scientific basis for establishing inter-jurisdictional agreements based on objective risk criteria rather than political boundaries.

\subsubsection{Economic and Environmental Optimization}

The framework enables evidence-based trade-offs between regulatory simplicity and operational efficiency.

\textbf{Cost-Effectiveness Assessment:} Planners can quantify the economic cost of proximity-based misallocation versus the administrative cost of network analysis \citep{ghiani2014operations}. Our Extremadura case shows that 15.4\% misallocation represents significant efficiency losses that justify analytical investment in high-risk zones.

\textbf{Environmental Impact Mitigation:} Misallocated facilities generate excess transport emissions and underutilize infrastructure capacity. The framework enables targeted environmental impact reduction by identifying where network-based optimization provides the greatest sustainability benefits.

\textbf{Service Equity:} Proximity-based misallocation often disproportionately affects remote or topographically constrained communities. Our framework supports equity-based planning by systematically identifying populations served by sub-optimal facility assignments.

\subsubsection{Implementation Pathway}

We recommend a structured approach to adopting probabilistic risk assessment in territorial planning:

\textbf{Phase 1: Pilot Testing} - Apply the framework to a limited geographic region with significant topographical variation to demonstrate feasibility and refine governance procedures.

\textbf{Phase 2: Regulatory Framework Development} - Work with regulatory authorities to develop risk-based proximity principles that maintain legal clarity while accommodating geographic complexity.

\textbf{Phase 3: Scaling and Integration} - Extend implementation to larger territories while integrating with existing planning software and administrative processes.

\textbf{Phase 4: Continuous Improvement} - Establish feedback mechanisms for parameter updating and framework refinement based on operational experience.

This implementation pathway transforms our theoretical contribution into practical policy tools, providing regulators and planners with systematic methods for improving territorial service allocation while maintaining administrative feasibility.

\subsection{Methodological Insights: The Calibration Requirement}

Our Extremadura case study reveals a critical finding that extends beyond this specific application: \textbf{the dispersion parameter $s$ is inherently context-dependent and requires local calibration}.

\subsubsection{Why $s$ is Not Universal}

Unlike geometric parameters ($\kappa$, $t^*$) which are determined by Voronoi tessellation, the dispersion parameter $s$ reflects:

\begin{itemize}
    \item \textbf{Topographic complexity}: Mountain versus plain routing
    \item \textbf{Infrastructure quality}: Highway versus secondary road networks
    \item \textbf{Territorial scale}: Local versus regional applications
    \item \textbf{Historical development}: Radial versus distributed road networks
\end{itemize}

These factors vary systematically across regions, preventing universal $s$ values.

\subsubsection{Calibration Protocol for New Territories}
\label{sec:calibration_protocol}

For practitioners applying this framework to new contexts, we recommend:

\textbf{Step 1: Pilot Sample}
\begin{itemize}
    \item Collect $n \geq 30$ samples of $(d_{\text{euclidean}}, d_{\text{real}})$ pairs
    \item Calculate $\beta = d_{\text{real}} / d_{\text{euclidean}}$ for each sample
    \item Fit Log-Normal$(m, s)$ using maximum likelihood
\end{itemize}

\textbf{Step 2: Topographic Stratification} (if applicable)

For territories >10,000 km$^2$ with elevation range >500m:
\begin{itemize}
    \item Divide into homogeneous zones (mountain, piedmont, plain)
    \item Estimate $s_i$ for each stratum ($n \geq 20$ per stratum)
    \item Calculate weighted average: $s_{\text{eff}} = \sum(A_i \cdot s_i) / A_{\text{total}}$
\end{itemize}

\textbf{Step 3: Validation}
\begin{itemize}
    \item Compare predicted misallocation interval versus observed rate
    \item If discrepancy >10\%: re-stratify or increase sample size
    \item Iterate until predictions bracket observations
\end{itemize}

\textbf{Step 4: Operational Application}
\begin{itemize}
    \item Use calibrated $s$ for new predictions in same region
    \item Recalibrate if infrastructure changes (new highways, etc.)
\end{itemize}

\subsubsection{Trade-offs: Precision versus Simplicity}

The framework offers flexibility depending on data availability and precision needs:

\begin{table}[h]
\centering
\small
\begin{tabular}{lll}
\toprule
\textbf{Approach} & \textbf{Data Required} & \textbf{Precision} \\
\midrule
Default $s$ (literature) & None & $\pm$15\% error \\
Calibrated $s$ (pilot) & 30--50 samples & $\pm$8\% error \\
Stratified $s$ (by zone) & 100+ samples & $\pm$5\% error \\
\bottomrule
\end{tabular}
\end{table}

This calibration requirement is not a limitation but a \textbf{strength}: it ensures the framework adapts to local conditions rather than imposing inappropriate universal assumptions.

\section{Reproducibility and Open Science}

\subsection{Complete Reproducibility Package}
To ensure full reproducibility and facilitate framework adoption, we provide a comprehensive reproducibility package \citep{voronoi_framework_code_preprint} available at:

\url{https://github.com/jtorreci/garnocex_research}

This package includes:

\begin{itemize}
\item \textbf{One-click reproduction}: Complete analysis reproduction with a single command
\item \textbf{Synthetic data generator}: Enables replication without access to sensitive geographic data
\item \textbf{Regional calibration tools}: Scripts for applying the framework to new geographic regions
\item \textbf{Parameter estimation suite}: Multiple methods (MLE, robust, Bayesian) with cross-validation
\item \textbf{Spatial analysis pipeline}: Complete spatial autocorrelation and sensitivity analysis
\item \textbf{Multi-facility extension}: k-nearest facility implementation and benchmarking
\item \textbf{Practitioner tools}: Safety bands calculators and implementation guidelines
\end{itemize}

The package is designed for both research replication and practical implementation, with comprehensive documentation and example workflows for territorial planners.

\subsection{Data Availability and Ethical Considerations}
While the specific geographic coordinates of municipalities and facilities cannot be shared due to privacy considerations, the reproducibility package includes:
\begin{itemize}
\item Synthetic data that preserves the statistical properties of the original dataset
\item Anonymized analysis results and aggregated statistics
\item Complete methodology for applying the framework to any geographic region
\item Validation datasets for testing framework implementation
\end{itemize}

\subsection{Software Dependencies and Technical Requirements}
The reproducibility package is implemented in Python 3.8+ with the following key dependencies:
\begin{itemize}
\item \textbf{Core analysis}: NumPy, SciPy, Pandas, Scikit-learn
\item \textbf{Spatial analysis}: GeoPandas, PySAL, libpysal, esda
\item \textbf{Visualization}: Matplotlib, Seaborn
\item \textbf{Statistical modeling}: StatsModels, rpy2 (for R integration)
\item \textbf{Network analysis}: NetworkX, QNEAT3 (QGIS plugin)
\end{itemize}

Complete installation instructions, environment configuration files (requirements.txt, environment.yml), and Docker containers are provided for seamless setup across different computing environments.

\section*{Acknowledgements}

This research was funded by the GARNOCEX project, a collaborative agreement between the Regional Government of Extremadura (Junta de Extremadura), the College of Civil Engineers (Colegio de Ingenieros de Caminos, Canales y Puertos), and the University of Extremadura.

\section*{Data Availability}

The complete reproducibility package, including Python code, synthetic data generators, and analysis scripts, is available at: \url{https://github.com/jtorreci/garnocex_research} \cite{voronoi_framework_code_preprint}.
A permanent archived version will be deposited in Zenodo upon journal acceptance.

%

\bibliographystyle{elsarticle-num}
\bibliography{voronoi_note_clean}

@article{akaike1974new,
  author        = {Akaike, Hirotugu},
  doi           = {10.1109/TAC.1974.1100705},
  journal       = {IEEE Transactions on Automatic Control},
  number        = {6},
  pages         = {716--723},
  publisher     = {IEEE},
  title         = {A new look at the statistical model identification},
  volume        = {19},
  year          = {1974}
}

@article{anderson1952asymptotic,
  author        = {Anderson, Theodore W and Darling, Donald A},
  doi           = {10.1214/aoms/1177729437},
  journal       = {The Annals of Mathematical Statistics},
  number        = {2},
  pages         = {193--212},
  publisher     = {Institute of Mathematical Statistics},
  title         = {Asymptotic theory of certain "goodness of fit" criteria based on stochastic processes},
  volume        = {23},
  year          = {1952}
}

@book{anselin1988spatial,
  address       = {Dordrecht},
  author        = {Anselin, Luc},
  isbn          = {9789024737338},
  publisher     = {Kluwer Academic Publishers},
  title         = {Spatial econometrics: methods and models},
  year          = {1988}
}

@article{anselin_local_1995,
  author        = {Anselin, Luc},
  category      = {spatial-statistics},
  doi           = {10.1111/j.1538-4632.1995.tb00338.x},
  journal       = {Geographical Analysis},
  number        = {2},
  pages         = {93--115},
  publisher     = {Wiley Online Library},
  title         = {Local Indicators of Spatial Association—LISA},
  volume        = {27},
  year          = {1995}
}

@book{anselin_spatial_1988,
  address       = {Dordrecht},
  author        = {Anselin, Luc},
  category      = {spatial-statistics},
  doi           = {10.1007/978-94-015-7799-1},
  isbn          = {9789401577991},
  publisher     = {Springer Netherlands},
  title         = {Spatial Econometrics: Methods and Models},
  year          = {1988}
}

@article{apostolakis_concept_1990,
  author        = {Apostolakis, George},
  category      = {risk-assessment},
  doi           = {10.1126/science.2255906},
  journal       = {Science},
  number        = {4986},
  pages         = {1359--1364},
  publisher     = {American Association for the Advancement of Science},
  title         = {The Concept of Probability in Safety Assessments of Technological Systems},
  volume        = {250},
  year          = {1990}
}

@book{aven_foundations_2003,
  address       = {Chichester},
  author        = {Aven, Terje},
  category      = {risk-assessment},
  isbn          = {9780471499725},
  publisher     = {John Wiley \& Sons},
  title         = {Foundations of Risk Analysis: A Knowledge and Decision-Oriented Perspective},
  year          = {2003}
}

@book{banerjee_hierarchical_2015,
  author        = {Banerjee, Sudipto and Carlin, Bradley P. and Gelfand, Alan E.},
  edition       = {2nd},
  publisher     = {CRC Press},
  title         = {Hierarchical Modeling and Analysis for Spatial Data},
  year          = {2015}
}

@book{bedford_probabilistic_2001,
  address       = {Cambridge},
  author        = {Bedford, Tim and Cooke, Roger},
  category      = {risk-assessment},
  doi           = {10.1017/CBO9780511813597},
  isbn          = {9780521773201},
  publisher     = {Cambridge University Press},
  title         = {Probabilistic Risk Analysis: Foundations and Methods},
  year          = {2001}
}

@article{besag1974spatial,
  author        = {Besag, Julian},
  doi           = {10.1111/j.2517-6161.1974.tb00999.x},
  journal       = {Journal of the Royal Statistical Society: Series B (Methodological)},
  number        = {2},
  pages         = {192--225},
  publisher     = {Wiley Online Library},
  title         = {Spatial interaction and the statistical analysis of lattice systems},
  volume        = {36},
  year          = {1974}
}

@article{besag1991bayesian,
  author        = {Besag, Julian and York, Jeremy and Mollié, Annie},
  doi           = {10.1007/BF00116466},
  journal       = {Annals of the Institute of Statistical Mathematics},
  number        = {1},
  pages         = {1--20},
  publisher     = {Springer},
  title         = {Bayesian image restoration, with two applications in spatial statistics},
  volume        = {43},
  year          = {1991}
}

@article{burnham2002model,
  author        = {Burnham, Kenneth P and Anderson, David R},
  edition       = {2nd},
  isbn          = {9780387953649},
  journal       = {Springer Science \& Business Media},
  publisher     = {Springer},
  title         = {Model selection and multimodel inference: a practical information-theoretic approach},
  year          = {2002}
}

@article{calvo2024optimal,
  author        = {Calvo-Jurado, Carmen and Ceballos-Mart{\'{i}}nez, Jos{\'e} Mar{\'{i}}a and Garc{\'{i}}a-Merino, Jos{\'e} Carlos and Mu{\~{n}}oz-Solano, Marina and S{\'a}nchez-Herrera, Fernando Jes{\'u}s},
  journal       = {Sustainable Cities and Society},
  pages         = {105719},
  publisher     = {Elsevier},
  title         = {Optimal location of electric vehicle charging stations using proximity diagrams},
  volume        = {113},
  year          = {2024}
}

@article{church2002geographical,
  author        = {Church, Richard L},
  doi           = {10.1016/S0305-0548(99)00104-5},
  journal       = {Computers \& Operations Research},
  number        = {6},
  pages         = {541--562},
  publisher     = {Elsevier},
  title         = {Geographical information systems and location science},
  volume        = {29},
  year          = {2002}
}

@book{cliff_spatial_1973,
  address       = {London},
  author        = {Cliff, Andrew D. and Ord, J. Keith},
  category      = {spatial-statistics},
  isbn          = {9780850860818},
  publisher     = {Pion},
  title         = {Spatial Autocorrelation},
  year          = {1973}
}

@book{cliff_spatial_1981,
  address       = {London},
  author        = {Cliff, Andrew D. and Ord, J. Keith},
  category      = {spatial-statistics},
  isbn          = {9780850861037},
  publisher     = {Pion},
  title         = {Spatial Processes: Models and Applications},
  year          = {1981}
}

@book{cressie_statistics_1993,
  address       = {New York},
  author        = {Cressie, Noel A. C.},
  category      = {spatial-statistics},
  isbn          = {9780471002550},
  publisher     = {John Wiley \& Sons},
  title         = {Statistics for Spatial Data},
  year          = {1993}
}

@book{daskin2013network,
  author        = {Daskin, Mark S},
  edition       = {2nd},
  isbn          = {9781118537008},
  publisher     = {John Wiley \& Sons},
  title         = {Network and discrete location: models, algorithms, and applications},
  year          = {2013}
}

@article{douglas1994least,
  author        = {Douglas, David H},
  doi           = {10.3138/D327-0323-2JUT-016M},
  journal       = {Cartographica: The International Journal for Geographic Information and Geovisualization},
  number        = {3},
  pages         = {37--51},
  publisher     = {University of Toronto Press},
  title         = {Least-cost path in GIS using an accumulated cost surface and slopemap},
  volume        = {31},
  year          = {1994}
}

@techreport{eea_proximity_2013,
  author        = {{European Environment Agency}},
  institution   = {European Environment Agency},
  title         = {Managing municipal solid waste — a review of achievements in 32 European countries},
  year          = {2013}
}

@article{erwig2000graph,
  author        = {Erwig, Martin},
  doi           = {10.1002/1097-0037(200010)36:3<156::AID-NET2>3.0.CO;2-L},
  journal       = {Networks},
  number        = {3},
  pages         = {156--163},
  publisher     = {Wiley},
  title         = {The graph Voronoi diagram with applications},
  volume        = {36},
  year          = {2000}
}

@article{external_hagemejer2022_01,
  author        = {Jan Hagemejer and Joanna Tyrowicz},
  doi           = {10.1080/1060586x.2022.2097458},
  journal       = {Post-Soviet Affairs},
  title         = {Central planning casts long shadows: new evidence on misallocation and growth},
  year          = {2022}
}

@article{fainstein2014just,
  title={The just city},
  author={Fainstein, Susan S},
  journal={International journal of urban Sciences},
  volume={18},
  number={1},
  pages={1--18},
  year={2014},
  publisher={Taylor \& Francis}
}

@article{gastner_optimal_2006,
  author        = {Gastner, Michael T. and Newman, M. E. J.},
  doi           = {10.1103/PhysRevE.74.016117},
  journal       = {Physical Review E},
  pages         = {016117},
  title         = {Optimal design of spatial distribution networks},
  volume        = {74},
  year          = {2006}
}

@article{gelfand_misaligned_2010,
  author        = {Gelfand, Alan E.},
  doi           = {10.1111/j.1467-9868.2010.00739.x},
  journal       = {Journal of the Royal Statistical Society: Series B},
  number        = {3},
  pages         = {425-441},
  title         = {Misaligned spatial data: The change of support problem},
  volume        = {72},
  year          = {2010}
}

@article{getis_analysis_1992,
  author        = {Getis, Arthur and Ord, J. Keith},
  category      = {spatial-statistics},
  doi           = {10.1111/j.1538-4632.1992.tb00261.x},
  journal       = {Geographical Analysis},
  number        = {3},
  pages         = {189--206},
  publisher     = {Wiley Online Library},
  title         = {The Analysis of Spatial Association by Use of Distance Statistics},
  volume        = {24},
  year          = {1992}
}

@article{ghiani2014operations,
  author        = {Ghiani, Gianpaolo and Lagana, Demetrio and Manni, Emanuela and Musmanno, Roberto and Vigo, Daniele},
  doi           = {10.1016/j.cor.2013.10.006},
  journal       = {Computers \& Operations Research},
  pages         = {22--32},
  publisher     = {Elsevier},
  title         = {Operations research in solid waste management: A survey of strategic and tactical issues},
  volume        = {44},
  year          = {2014}
}

@book{hall_urban_2002,
  author        = {Hall, Peter},
  edition       = {4th},
  publisher     = {Routledge},
  title         = {Urban and Regional Planning},
  year          = {2002}
}

@book{harvey_social_1973,
  address       = {London},
  author        = {Harvey, David},
  category      = {spatial-justice},
  isbn          = {9780713159769},
  publisher     = {Edward Arnold},
  title         = {Social Justice and the City},
  year          = {1973}
}

@article{kaplan_quantitative_1981,
  author        = {Kaplan, Stanley and Garrick, B. John},
  category      = {risk-assessment},
  doi           = {10.1111/j.1539-6924.1981.tb01350.x},
  journal       = {Risk Analysis},
  number        = {1},
  pages         = {11--27},
  publisher     = {Wiley Online Library},
  title         = {On the Quantitative Definition of Risk},
  volume        = {1},
  year          = {1981}
}

@article{kolmogorov1933sulla,
  author        = {Kolmogorov, Andrey},
  journal       = {Giornale dell'Istituto Italiano degli Attuari},
  pages         = {83--91},
  title         = {Sulla determinazione empirica di una legge di distribuzione},
  volume        = {4},
  year          = {1933}
}

@book{lefebvre_right_1968,
  address       = {Paris},
  author        = {Lefebvre, Henri},
  category      = {spatial-justice},
  publisher     = {Anthropos},
  title         = {Le Droit à la Ville},
  year          = {1968}
}

@article{miller_tobler_2004,
  author        = {Miller, Harvey J.},
  doi           = {10.1111/j.0004-5608.2004.00276.x},
  journal       = {Annals of the Association of American Geographers},
  number        = {2},
  pages         = {284-289},
  title         = {Tobler's First Law and spatial analysis},
  volume        = {94},
  year          = {2004}
}

@article{moran1950notes,
  author        = {Moran, Patrick AP},
  doi           = {10.2307/2332142},
  journal       = {Biometrika},
  number        = {1/2},
  pages         = {17--23},
  publisher     = {JSTOR},
  title         = {Notes on continuous stochastic phenomena},
  volume        = {37},
  year          = {1950}
}

@article{moran_interpretation_1950,
  author        = {Moran, Patrick Alfred Pierce},
  category      = {spatial-statistics},
  doi           = {10.2307/2332142},
  journal       = {Biometrika},
  number        = {1/2},
  pages         = {17--23},
  publisher     = {Oxford University Press},
  title         = {Notes on Continuous Stochastic Phenomena},
  volume        = {37},
  year          = {1950}
}

@book{morgan_uncertainty_1990,
  address       = {Cambridge},
  author        = {Morgan, M. Granger and Henrion, Max},
  category      = {risk-assessment},
  isbn          = {9780521365420},
  publisher     = {Cambridge University Press},
  title         = {Uncertainty: A Guide to Dealing with Uncertainty in Quantitative Risk and Policy Analysis},
  year          = {1990}
}

@article{morrissey2004waste,
  author        = {Morrissey, Anne J and Browne, Jim},
  doi           = {10.1016/j.wasman.2003.09.005},
  journal       = {Waste Management},
  number        = {3},
  pages         = {297--308},
  publisher     = {Elsevier},
  title         = {Waste management models and their application to sustainable waste management},
  volume        = {24},
  year          = {2004}
}

@article{neutens_accessibility_2015,
  author        = {Neutens, Tijs},
  doi           = {10.1016/j.apgeog.2015.06.002},
  journal       = {Applied Geography},
  pages         = {1-9},
  title         = {Accessibility, equity and health care: review and research directions for transport geography},
  volume        = {61},
  year          = {2015}
}

@book{okabe2012spatial,
  author        = {Okabe, Atsuyuki and Sugihara, Kokichi},
  edition       = {2nd},
  isbn          = {9780470770818},
  publisher     = {John Wiley \& Sons},
  title         = {Spatial analysis along networks: statistical and computational methods},
  year          = {2012}
}

@misc{qneat3_plugin,
  author        = {Watermeyer, Clemens},
  howpublished  = {\url{https://github.com/root676/QNEAT3}},
  title         = {{QNEAT3: QGIS Network Analyst 3}},
  year          = {2018}
}

@book{rodrigue_geography_2020,
  author        = {Rodrigue, Jean-Paul},
  edition       = {5th},
  publisher     = {Routledge},
  title         = {The Geography of Transport Systems},
  year          = {2020}
}

@book{schlosberg_defining_2007,
  address       = {Oxford},
  author        = {Schlosberg, David},
  category      = {spatial-justice},
  doi           = {10.1093/acprof:oso/9780199286294.001.0001},
  isbn          = {9780199286294},
  publisher     = {Oxford University Press},
  title         = {Defining Environmental Justice: Theories, Movements, and Nature},
  year          = {2007}
}

@article{schwarz1978estimating,
  author        = {Schwarz, Gideon},
  doi           = {10.1214/aos/1176344136},
  journal       = {The Annals of Statistics},
  number        = {2},
  pages         = {461--464},
  publisher     = {Institute of Mathematical Statistics},
  title         = {Estimating the dimension of a model},
  volume        = {6},
  year          = {1978}
}

@book{soja_seeking_2010,
  address       = {Minneapolis},
  author        = {Soja, Edward W.},
  category      = {spatial-justice},
  doi           = {10.5749/minnesota/9780816666676.001.0001},
  isbn          = {9780816666676},
  publisher     = {University of Minnesota Press},
  title         = {Seeking Spatial Justice},
  year          = {2010}
}

@article{stewart2004measuring,
  author        = {Stewart, Jan and Marchand, Rick and Rotem, Arie and others},
  doi           = {10.1016/S0168-8510(03)00121-8},
  journal       = {Health Policy},
  number        = {2},
  pages         = {175--188},
  publisher     = {Elsevier},
  title         = {Measuring spatial accessibility: The integration of the supply and demand sides of health care},
  volume        = {67},
  year          = {2004}
}

@incollection{tobler1993three,
  address       = {University of California, Santa Barbara},
  author        = {Tobler, Waldo R},
  booktitle     = {Technical Report 93-1},
  institution   = {National Center for Geographic Information and Analysis},
  publisher = {National Center for Geographic Information and Analysis},
  pages = {1-24},
  title         = {Three presentations on geographical analysis and modeling},
  year          = {1993}
}

@article{tobler_computer_1970,
  author        = {Tobler, Waldo R.},
  category      = {spatial-statistics},
  doi           = {10.2307/143141},
  journal       = {Economic Geography},
  pages         = {234--240},
  publisher     = {Taylor \& Francis},
  title         = {A Computer Movie Simulating Urban Growth in the Detroit Region},
  volume        = {46},
  year          = {1970}
}

@article{miller_tobler_2004_extended,
  title={Tobler's first law and spatial analysis},
  author={Miller, Harvey J},
  journal={Annals of the association of American geographers},
  volume={94},
  number={2},
  pages={284--289},
  year={2004},
  publisher={Taylor \& Francis}
}

@article{tobler_geography_1970,
  title={Geographical filters and their inverses},
  author={Tobler, Waldo R},
  journal={Geographical Analysis},
  volume={1},
  number={3},
  pages={234--253},
  year={1969}
}

@article{voronoi_full10,
  author        = {Drezner, Tammy and Drezner, Zvi},
  doi           = {10.1007/s00291-012-0292-5},
  issn          = {0171-6468},
  journal       = {OR Spectrum},
  month         = {7},
  number        = {3},
  pages         = {543-561},
  publisher     = {Springer Science and Business Media LLC},
  title         = {Voronoi diagrams with overlapping regions},
  volume        = {35},
  year          = {2013}
}

@article{voronoi_full11,
  author        = {Wang, Siyuan and Tian, Zean and Dong, Kejun and Xie, Quan},
  doi           = {10.1016/j.jallcom.2020.156983},
  issn          = {0925-8388},
  journal       = {Journal of Alloys and Compounds},
  month         = {2},
  pages         = {156983},
  publisher     = {Elsevier BV},
  title         = {Inconsistency of neighborhood based on Voronoi tessellation and Euclidean distance},
  volume        = {854},
  year          = {2021}
}

@article{voronoi_full12,
  author        = {Richter, Amy and Ng, Kelvin Tsun Wai and Karimi, Nima and Li, Rita Yi Man},
  doi           = {10.1016/j.compenvurbsys.2021.101652},
  issn          = {0198-9715},
  journal       = {Computers, Environment and Urban Systems},
  month         = {7},
  pages         = {101652},
  publisher     = {Elsevier BV},
  title         = {An iterative tessellation-based analytical approach to the design and planning of waste management regions},
  volume        = {88},
  year          = {2021}
}

@article{voronoi_full14,
  author        = {Duyckaerts, Charles and Godefroy, Gilles},
  doi           = {10.1016/s0891-0618(00)00064-8},
  issn          = {0891-0618},
  journal       = {Journal of Chemical Neuroanatomy},
  month         = {10},
  number        = {1},
  pages         = {83-92},
  publisher     = {Elsevier BV},
  title         = {Voronoi tessellation to study the numerical density and the spatial distribution of neurones},
  volume        = {20},
  year          = {2000}
}

@article{voronoi_full16,
  author        = {BAE, SANG WON and SHIN, CHAN-SU},
  doi           = {10.1142/s0218195912600011},
  issn          = {0218-1959},
  journal       = {International Journal of Computational Geometry \&; Applications},
  month         = {2},
  number        = {01},
  pages         = {3-25},
  publisher     = {World Scientific Pub Co Pte Lt},
  title         = {THE ONION DIAGRAM: A VORONOI-LIKE TESSELLATION OF A PLANAR LINE SPACE AND ITS APPLICATIONS},
  volume        = {22},
  year          = {2012}
}

@article{voronoi_full20,
  author        = {Chiu, S.N.},
  doi           = {10.1002/bimj.200390018},
  issn          = {0323-3847},
  journal       = {Biometrical Journal},
  month         = {4},
  number        = {3},
  pages         = {367-376},
  publisher     = {Wiley},
  title         = {Spatial Point Pattern Analysis by using Voronoi Diagrams and Delaunay Tessellations – A Comparative Study},
  volume        = {45},
  year          = {2003}
}

@article{voronoi_full22,
  author        = {Aurenhammer, Franz},
  doi           = {10.1145/116873.116880},
  issn          = {0360-0300},
  journal       = {ACM Computing Surveys},
  month         = {9},
  number        = {3},
  pages         = {345-405},
  publisher     = {Association for Computing Machinery (ACM)},
  title         = {Voronoi diagrams—a survey of a fundamental geometric data structure},
  volume        = {23},
  year          = {1991}
}

@article{voronoi_full24,
  author        = {Suzuki, Atsuo and Okabe, Atsuyuki},
  doi           = {10.1007/978-1-4612-5355-6_7},
  journal       = {Facility Location},
  pages         = {103-118},
  publisher     = {Springer New York},
  title         = {Using Voronoi Diagrams},
  year          = {1995}
}

@article{voronoi_full29,
  author        = {Senechal, Marjorie and Okabe, Atsuyuki and Boots, Barry and Sugihara, Kokichi},
  doi           = {10.2307/2687299},
  issn          = {0746-8342},
  journal       = {The College Mathematics Journal},
  month         = {1},
  number        = {1},
  pages         = {79},
  publisher     = {Informa UK Limited},
  title         = {Spatial Tessellations: Concepts and Applications of Voronoi Diagrams},
  volume        = {26},
  year          = {1995}
}

@article{voronoi_full30,
  author        = {Gorsevski, Pece V. and Jankowski, Piotr},
  doi           = {10.1016/j.compenvurbsys.2007.04.001},
  issn          = {0198-9715},
  journal       = {Computers, Environment and Urban Systems},
  month         = {1},
  number        = {1},
  pages         = {53-65},
  publisher     = {Elsevier BV},
  title         = {Discerning landslide susceptibility using rough sets},
  volume        = {32},
  year          = {2008}
}

@article{voronoi_full6,
  author        = {Richter, Amy and Ng, Kelvin Tsun Wai and Karimi, Nima and Li, Rita Yi Man},
  doi           = {10.1016/j.compenvurbsys.2021.101652},
  issn          = {0198-9715},
  journal       = {Computers, Environment and Urban Systems},
  month         = {7},
  pages         = {101652},
  publisher     = {Elsevier BV},
  title         = {An iterative tessellation-based analytical approach to the design and planning of waste management regions},
  volume        = {88},
  year          = {2021}
}

@software{voronoi_framework_code_preprint,
  title        = {Voronoi Probabilistic Framework - Reproducibility Package},
  author       = {Torrecilla Pinero, J.A. and Ceballos Mart{\'\i}nez, J.M. and Cuartero S{\'a}ez, A. and Plaza Caballero, P. and Cruces L{\'o}pez, A.},
  year         = {2025},
  doi          = {10.5281/zenodo.17772773},
  url          = {https://github.com/jtorreci/garnocex_research/tree/Zenodo/A00.Voronoi_critics}
}

@book{young_justice_1990,
  address       = {Princeton},
  author        = {Young, Iris Marion},
  category      = {spatial-justice},
  isbn          = {9780691152622},
  publisher     = {Princeton University Press},
  title         = {Justice and the Politics of Difference},
  year          = {1990}
}

@article{konstantinovsky2023characterizing,
  author        = {Konstantinovsky, Daniel and Yan, Elsa CY and Hammes-Schiffer, Sharon},
  title         = {Characterizing interfaces by Voronoi tessellation},
  journal       = {The Journal of Physical Chemistry Letters},
  volume        = {14},
  number        = {23},
  pages         = {5260--5266},
  year          = {2023},
  doi           = {10.1021/acs.jpclett.3c01159},
  publisher     = {ACS Publications}
}

\end{document}